\documentclass[letterpaper,11pt]{article}
\usepackage{amsmath,amssymb,amsthm,amsfonts}
\usepackage[top=8pc,bottom=8pc,left=8pc,right=8pc]{geometry}

\usepackage{graphicx}
\usepackage{epstopdf}
\usepackage{subcaption}
\usepackage{multirow}

\usepackage{times}
\usepackage{bm}

\RequirePackage[hyphens]{url}
\RequirePackage[colorlinks,citecolor=blue,urlcolor=blue]{hyperref}
\usepackage[authoryear]{natbib}

\usepackage[plain,noend]{algorithm2e}

\usepackage{array}
\newcolumntype{L}[1]{>{\raggedright\let\newline\\\arraybackslash\hspace{0pt}}m{#1}}
\newcolumntype{C}[1]{>{\centering\let\newline\\\arraybackslash\hspace{0pt}}m{#1}}
\newcolumntype{R}[1]{>{\raggedleft\let\newline\\\arraybackslash\hspace{0pt}}m{#1}}

 \newcommand{\beginsupplement}{%
        \setcounter{equation}{0}
				\setcounter{page}{0}
				\setcounter{table}{0}
				\setcounter{section}{0}
				\setcounter{figure}{0}

			\renewcommand{\theequation}{S.\arabic{equation}}
			\renewcommand{\thesection}{S.\arabic{section}}
			\renewcommand{\thesubsection}{S.\arabic{section}.\arabic{subsection}}
			\renewcommand{\thepage}{S.\arabic{page}}
			\renewcommand{\thetable}{S.\arabic{table}}
			\renewcommand{\thefigure}{S.\arabic{figure}}
}
  
\newcommand{\I}{{\bf I}}

\newcommand{\Nor}{{\cal N}}  

\renewcommand{\S}{{\bf S}}

\newcommand{\ben}{\begin{enumerate}}
\newcommand{\een}{\end{enumerate}}
\newcommand{\beq}{\begin{equation}}
\newcommand{\eeq}{\end{equation}}

\newcommand{\half}{\frac{1}{2}}

\newcommand{\vectornorm}[1]{\left|\left|#1\right|\right|}

\newcommand{\Norm}{\mathcal{N}}

\newtheorem{theorem}{Theorem}
\numberwithin{theorem}{section}

\newtheorem{Def}{Definition}
\numberwithin{Def}{section}

\numberwithin{remark}{section}

\newtheorem{proposition}{Proposition}
\numberwithin{proposition}{section}

\numberwithin{lemma}{section}

\numberwithin{Cor}{section}

\renewcommand {\S}{\textsection} 

\title{Default Bayesian analysis with global-local shrinkage priors}

\author{Anindya Bhadra  \footnote{{\em Address:} 250 N. University St., West Lafayette, IN 47907, email: bhadra@purdue.edu.} \\Purdue University
\and Jyotishka Datta  \footnote{{\em Address:} Box 90251, Durham, NC 27708, email: jd298@stat.duke.edu.}\\ Duke University and SAMSI\\
\and Nicholas G. Polson \footnote{{\em Address:} 5807 S. Woodlawn Ave., Chicago, IL 60637, email: ngp@chicagobooth.edu.}  \ \ and Brandon Willard \footnote{{\em Address:} 5807 S. Woodlawn Ave., Chicago, IL 60637, email: bwillard@uchicago.edu.} \\The University of Chicago Booth School of Business}

\date{\vspace{-0.1cm} \today}
\IfFileExists{upquote.sty}{\usepackage{upquote}}{}
\begin{document}
\maketitle

\maketitle
\begin{abstract}
\noindent We provide a framework for assessing the default nature of a prior distribution using the property of regular variation, which we study for global-local shrinkage priors. In particular, we demonstrate the horseshoe priors, originally designed to handle sparsity, also possess regular variation and thus are appropriate for default Bayesian analysis. To illustrate our methodology, we solve a problem of non-informative priors due to \cite{efron1973discussion}, who showed standard flat non-informative priors in high-dimensional normal means model can be highly informative for nonlinear parameters of interest. We consider four such problems and show global-local shrinkage priors such as the horseshoe and horseshoe+  perform as \cite{efron1973discussion} requires in each case. We find the reason for this lies in the ability of the global-local shrinkage priors to separate a low-dimensional signal embedded in high-dimensional noise, even for nonlinear functions. 
\end{abstract}


\section{Introduction}\label{sec:intro}
\subsection{Default Bayesian analysis in multi-parameter problems}
In the discussion to \citet{dawid1973marginalization}, \citet{efron1973discussion} observed that the traditional default prior for the normal means could actually be highly informative for a low-dimensional nonlinear function of the means. \citet{efron1978controversies} further observes, ``the trouble with the multi-parameter estimation problem is not that it is harder than estimating a single parameter.  It is easier, in the sense that dealing with many problems simultaneously can give extra information not otherwise available. The trouble lies  in finding and using the extra information.'' Bayes' rule is clear on how to perform inference for a multivariate mean $ \theta=(\theta_1 , \ldots , \theta_p)$, as the posterior density is $ p( \theta \mid y ) \propto p( y \mid \theta ) p(\theta)$. The problem is in specifying $p(\theta)$. 

Consider the classic normal means model $(y_i \mid \theta_i) \sim \Nor (\theta_i, 1)$ independently for $i = 1, \ldots, p$. A common statistical inference problem occurs when there is a high-dimensional parameter vector $\theta$ but the parameter of interest $\psi$ is a low-dimensional nonlinear function. In the absence of relevant prior information, \citet{Jeffreys1961} argues ``if we have no information relevant to the actual value of the parameter, the probability must be chosen so as to express the fact that we have none.'' Jeffreys' non-informative prior, $p(\theta) \propto |I(\theta)|^{1/2}$, where $I(\theta)$ is the Fisher information,  is based on preservation of invariance under transformation, although for the normal means problem, he proposed a departure by using $\pi(\mu,\sigma) \propto 1/\sigma$. Unfortunately, Jeffreys' prior is often improper. \cite{dawid1973marginalization} argue that the posterior distributions for the improper priors have an unBayesian property, the result of the so-called marginalization paradox, which can be alleviated by using proper priors. 

\cite{efron1973discussion} illustrates that when the parameter of interest is a nonlinear function of normal means, the posterior could lead away from a non-informative answer even when using a proper, and yet flat prior on $\theta$, and argues that the flat priors only make sense for group-transformation models. However, group transformation encompasses only a limited class of models, and in many cases the parameter of interest could be a complicated nonlinear quantity beyond the reach of the flat priors. \cite{bernardo1979reference} introduced the important notion of a parameter of interest where one re-parameterizes $ \theta=(\psi, \zeta)$ where $\psi$ is a parameter of interest and  $\zeta$ is a vector of nuisance parameters that requires marginalization.

These re-parameterizations have little effect on simulation such as Markov chain Monte Carlo as one simply transforms the simulated draws from one parameterization to another. The issue that arises, however, is in specifying the default priors for inference. Transforming a prior, $p(\theta)$,  from a high-dimensional space, to a low-dimensional, generally nonlinear parameter of interest, $p(\psi)$, involves a Jacobian or density transform that can take a seemingly non-informative prior into a highly informative one, leaving the researcher with a posterior distribution given the data that does not represent beliefs associated with non-informativity. Jeffreys' prior suffers directly from these issues and the growth of reference priors \citep{bernardo1979reference, berger1992development} was designed directly to handle the issue of non-informative inference for a parameter of interest while marginalizing out high-dimensional nuisance parameters. The problem of specifying a non-informative prior for a parameter of interest $ \psi $ then reduces to the properties of the marginal posterior of $\psi$ given $y$. The reference priors of \cite{bernardo1979reference} are constructed to not suffer from marginalization paradoxes. However, these priors are typically not integrable, which can lead to problems in more complicated settings such as hierarchical models \citep{gelman2006prior}. 

\citet{efron1973discussion} posed a number of challenges for a Bayesian framework for default prior analysis, including the sum of squares and maximum of normal means. Specifically, he noted that priors such as independent $N(0,A)$ prior for each $\theta_i$ where $ A \rightarrow \infty$  can be highly informative for non-linearly transformed low-dimensional parameters of interest, such as, $\psi = \sum_{i=1}^{p} \theta_i^2$.  A similar issue arises with estimating the variance $\sigma^2$ in the model $(y \mid \theta, \sigma^2) \sim \Nor (\theta, \sigma^2\I)$ under the inverse gamma prior $\sigma^2 \sim IG ( \epsilon, \epsilon ) $ as $ \epsilon \rightarrow 0$.  \citet{gelman2006prior} proposed a $C^+(0,2{\cdot}5)$ prior for the scale parameter in a hierarchical model, where $\sigma \sim C^+(0,\tau)$ denotes a half-Cauchy distributed random variable, with density $p(\sigma \mid \tau) = 2/\{\pi \tau(1 + (\sigma/\tau)^2\}$, $\sigma > 0, \tau > 0$. 

We argue here that the class of global-local shrinkage priors \citep{polson2012local} are good candidates for default priors. This class includes the horseshoe prior \citep{carvalho2010horseshoe} and the horseshoe+ prior introduced in an unpublished 2015 technical report by the authors of the current article. They are also particularly effective when the underlying parameter vector is sparse and the parameter of interest is a low-dimensional, possibly nonlinear function. \cite{gelman2006prior} already hinted at the default nature of half-Cauchy priors in hierarchical models.  The question we consider is:  what is the special property that distinguishes the half-Cauchy prior from other so-called flat priors? We will argue that the property of regular variation is key to a non-informative analysis. We study this property for the class of global-local shrinkage priors and show, via the use of dual densities, how to construct such priors. 


While the use of heavy-tailed robust priors for non-informative inference has a long history \citep{dawid1973marginalization, o1979outliers,o1990outliers}, our new insight is that it is important to use sparse-robust priors as well, that is, ones that offer heavy shrinkage towards zero. For example, in the ratio of means, also known as the Fieller--Creasy problem \citep{fieller1940biological,creasy1954limits}, where $\psi= \theta_1 / \theta_2 $, one needs a prior that will let the data speak for themselves at the origin of $\theta_2$. In the case where the true parameter values are $ \theta_1 = \theta_2 = 0$, the likelihood ought to be uninformative and we should have a posterior with a similar spread as the prior. 



\subsection{Global-local shrinkage} \label{sec:gl}
We define the following class of priors on $\theta_i$ as global-local shrinkage priors
$$
(\theta_i \mid \lambda_i, \tau) \sim \Nor(0, \lambda_i^2), (\lambda_i \mid \tau)  \sim p (\lambda_i \mid \tau), \tau \sim p(\tau), \quad \lambda_i > 0, \tau > 0, 
$$
where $p(\lambda_i \mid \tau)$ has a tail that decays slower than an exponential rate. 


Our approach to the marginal parameters of interest problems will be based on the use of global-local shrinkage priors \citep{polson2012local, polson2012half}. In particular, we focus on the horseshoe and horseshoe+ priors where the hierarchical models are specified as follows for the horseshoe,
\begin{align*}
  ( \theta_i \mid \lambda_i, \tau ) &\sim \Nor \left ( 0 , \lambda_i^2 \right ), ( \lambda_i \mid \tau ) \sim C^+ \left ( 0 , \tau \right ), \tau  \sim C^+ \left ( 0 , 1 \right ), \lambda_i, \tau > 0,
\end{align*}
and for the horseshoe+,
\begin{align*}
( \theta_i \mid \lambda_i, \eta_i, \tau ) &\sim \Nor \left(0 , \lambda_i^2 \right), (\lambda_i \mid \eta_i,\tau) \sim C^+ \left( 0 , \tau \eta_i \right), \\
	\eta_i & \sim C^+ \left( 0 , 1 \right), \tau  \sim C^+ \left ( 0 , 1 \right ), \lambda_i, \eta_i, \tau > 0. 
 \end{align*}
A feature of the horseshoe and horseshoe+ priors is that they possess both tail-robustness and sparse-robustness properties, in other words, an infinite spike at the origin and as heavy a tail as possible that still ensures integrability. 

It is useful to view these priors by transforming to a shrinkage scale with
$\kappa_i = 1/(1 + \lambda_i^2 \tau^2)$. This transformation yields
the densities of global-local shrinkage priors, 
$$ 
  p(\theta_i \mid \tau) = \int_0^1 p(\theta_i \mid \kappa_i, \tau) p(\kappa_i \mid \tau) d\kappa_i, 
  \; p(\theta_i \mid \kappa_i, \tau) \sim \mathcal{N} \left(0, \frac{1-\kappa_i}{\kappa_i} \right).
$$ 
Here $ \kappa_i \in [0,1]$ is a component-specific shrinkage weight and $\tau$ is a global parameter. 

In the case of a sparse $\theta$, only a few components are away from zero. \citet{west84} notes that the solution in this case is not to use a spherically symmetric heavy-tailed prior, such as a multivariate $t$. Indeed, for most of the components, the prior belief matches the observations, differing only in a few coordinates. Thus, the solution lies in putting component-specific priors on $\theta$. This is also the basic idea in global-local shrinkage priors.  The global $\tau$ provides shrinkage towards zero for all components and indeed, the marginal prior on $\theta_i$ has unbounded spike at zero. The local, or component-specific $\lambda_i$ terms leave the large signals un-shrunk. The further half-Cauchy priors on $\lambda_i$ and $\tau$ allow these terms to be learned from the data. There are several optimality results available for the horseshoe and horseshoe+ priors \citep[][for the horseshoe; and a 2015 technical report by the current authors for the horseshoe+]{carvalho2010horseshoe,datta2013asymptotic,van2014horseshoe}. However, all of these results are concerned with the estimation of the multivariate normal mean itself, rather than its low-dimensional functions, which is the focus of the current work. We now discuss a key property that allows the global-local priors to perform well in the problems we consider: that of regular variation.

\section{Regular Variation and Non-informative Priors}\label{sec:heavy}
\subsection{Regular variation and Bayes' rule}\label{sec:rv-bayes}
We consider Bayes' rule for the parameter of interest $\psi$
$$
p( \psi \mid y ) = \frac{ L(\psi) p(\psi)}{p(y)}, \; \text{where} \; \; L(\psi) = \int p( y \mid \xi , \psi ) p( \xi \mid \psi ) d \xi, 
$$
is the marginal likelihood.  We need a prior $p(\theta_1, \ldots , \theta_p)$ that has a conditional $p(\xi \mid \psi) $ that leads to a well-behaved marginal likelihood and we need the prior $p(\psi)$ to have regular variation relative to this marginal likelihood. A positive, measurable function $f$ is called regularly varying at infinity with index $\alpha$ if $f$ is defined on some neighborhood $[x_0,\infty)$ and $\lim_{x \to \infty} f(tx)/f(x) \to t^{\alpha}$ for all $t>0$. The function $f$ is called slowly varying if $\alpha = 0$. A random variable $Y$ with distribution $F$ is defined to have a regularly varying upper tail if its survival function $\bar{F}$ is regularly varying at infinity. The regularly varying functions also satisfy a number of Tauberian theorems \citep{bingham1989regular}. A useful outcome of these theorems is that they are closed under most common many-to-one functions, for example, sums of regularly varying random variables are also regularly varying. The last property also characterizes a wider class of heavy-tailed distributions called sub-exponential distribution, which has also been linked to Bayesian robustness by \cite{andrade2011bayesian-b}. 

A key idea in Bayesian robustness modeling is the use of heavy-tailed distributions to resolve the conflict between different sources of information, namely the prior and outliers in data, that light-tailed distributions such as Gaussian fail to account for. The idea goes back to \citet{de1961bayesian}, \citet{lindley1968choice}, and \citet{dawid1973marginalization}. \citet{lindley1968choice} showed, by means of an approximate analysis, that if the likelihood is $y \mid \theta \sim \Nor(\theta, \sigma^2)$, then the posterior mean for $\theta$ given $y$ would not be too far from the observation $y$, if one uses a Student's $t$ prior instead of a Gaussian prior on $\theta$. \citet{dawid1973posterior} generalized the result for a pure location family and provided conditions that the prior ``should not wag its tail too vigorously," in order to achieve outlier rejection. The connection between heavy tails and Bayesian robustness has been explored in depth in the subsequent literature by \citet{o1979outliers,o1990outliers,andrade2006bayesian,andrade2011bayesian} and \citet{o2012bayesian}. In particular, \citet{andrade2006bayesian,andrade2011bayesian} established easily verifiable properties for conflict resolution through regular variation of the prior for location-scale families. 

A remarkable result in the theory of regular variation is due to \cite{barndorff1982normal} and we show its use in Bayesian robustness. Their theorem shows that for any mixing distribution with slowly varying tail behavior, the corresponding normal variance mixture will also have slowly varying tails. In fact, the argument can be extended to any normal mean-variance mixtures by identifying a mean-variance mixture as an exponentially tilted version of the pure variance-mixture. This result shows that putting a slowly varying prior on the local shrinkage parameters $\Lambda = (\lambda_1^2, \ldots, \lambda_p^2)$ is a straightforward way of constructing slowly varying priors for the parameter of interest $\theta$. 

\begin{theorem}\label{th:barndorff} \citep{barndorff1982normal}.
Let $p(\cdot)$ be a normal scale mixture $p(\theta) = \int (2\pi v)^{-1/2} e^{-\theta^2/2v} dF(v)$ and suppose that the mixing density $f(\cdot)$ satisfies 
\beq 
f(v) = e^{-\psi^+v} v^{\alpha-1} L(v), \; v \to \infty, \nonumber
\eeq
where $\psi^+ = \sup\{ s \in \Re: \phi(s) = \int e^{su} F(du) < \infty \}$, and $L(\cdot)$ is slowly-varying. Then, 
\[
  p(\theta) \sim 
  \begin{cases} 
    (2\pi)^{-\half} 2^{\half-\alpha} \Gamma(\half-\alpha) |\theta|^{2\alpha-1} L(\theta^2) , \; \theta \to \infty, \psi^+ = 0 \\
    (2\psi^+)^{-\half \alpha} L(|\theta|) |\theta|^{\alpha-1} \exp\{-(2 \psi^+)|\theta|\}, \; \theta \to \infty, \psi^+ > 0.
  \end{cases}
\]
\end{theorem}

The results of \citet{barndorff1982normal}, together with Tweedie's formula for the exponential family \citep{efron2011tweedie} can be used for checking posterior robustness for scale mixtures of normals. Results on regular variation are reviewed in the Supplementary Material. 

Unlike the half-Cauchy prior, the normal and Laplace prior do not work well for these problems as they lack regular variation. A more subtle problem occurs for the inverse-gamma $(\epsilon, \epsilon)$ prior for the variance term $\nu$. The inverse-gamma prior has regular variation at the right tail but not at the origin as the $\exp(-\epsilon/\nu)$ term dominates $\nu^{-(1+\epsilon)}$ term near the origin, which leads to shrinkage away from the origin. 

We argue that the four classical one-dimensional problems introduced in \S \ref{sec:horseshoe} are examples of nice many-to-one functions that preserve the regular varying tails in the prior. It can be then argued that the heavy-tailed normal scale mixture priors such as the horseshoe and horseshoe+ will outperform priors for scale parameters with rapidly varying tails that make them unsuitable for robustness or conflict resolution. 

\subsection{Regular variation as a measure of flatness}

It has been argued that flat priors only make sense in group-transformation models \citep{efron1973discussion}. However, only a few models fall into that class, and often the goal is inference of complicated non-linear parameters of interest. As Bernardo points out in a conversation with \citet{irony1997non}, flat priors are usually very informative even when they are proper, and Jeffreys' prior works for regular, one-parameter problems but leads to marginalization paradoxes that both reference priors and the half-Cauchy priors seem to avoid for these multi-parameter problems. The heavy-tails of the half-Cauchy priors can be interpreted as a translation of vague prior beliefs \citep{rubio2014inference}. The usual non-informative priors become highly informative when one transforms a high-dimensional parameter into a low-dimensional parameter. 
However, regular variation, due to closure properties under a wide variety of transformations,  preserves the flatness of the priors even in the context of transformed parameters of interest. 

\subsection{Regular variation and nonlinear functionals}


Assume $\tau=1$ without loss of generality. The global-local prior \citep{polson2012local} has the density of $\lambda_i$ of the form $p(\lambda_i^2) = K (\lambda_i^2)^{-a-1} L(\lambda_i)$ for some $a>0$. \cite{polson2010shrink} and \cite{ghosh2015asymptotic} show that such priors achieve tail-robustness as well as asymptotic Bayes optimality when $L(\cdot)$ is a slowly varying function. The marginal prior on $\theta_i$ is
\begin{equation*}
p(\theta_i ) \propto \int (2\pi\lambda_i^2)^{-1/2} \exp\left(-\frac{\theta_i^2}{2 \lambda_i^2}\right) \lambda_i^{-a-1} L(\lambda_i^2) d (\lambda_i^2), \lambda_i >0, a \geq 0.
\end{equation*}
An application of Theorem \ref{th:barndorff} gives 
$$
p(\theta_i) \propto \left( \theta_i^2 \right)^{-a-1} L(\theta_i^2),  |\theta_i | \to \infty,
$$
which shows $\theta_i$ has a polynomially decaying density and hence a regularly varying right tail. Recall that the likelihood is normal and we have $[y_i \mid \theta_i] \sim \Nor(\theta_i,1)$. Thus, Result 2 of \citet{dawid1973posterior} implies $ [\theta_i \mid y_i] \rightsquigarrow \Nor(y_i, 1) $ as $|y_i | \to \infty$. 
Hence the posterior of $\theta_i$ corresponding to a large $y_i$ is centered at the observation $y_i$, with the same standard deviation as the data distribution. Thus, the posterior estimate of such a  $\theta_i$ will be the same as pure likelihood-based estimate and prior information is discounted. If one defines low dimensional functions of $\theta$ such as $\psi = \sum_{i=1}^{p } \theta_i^2, \psi = \max \theta_i, \psi = \theta_1\theta_2, \psi = \theta_1/\theta_2$, by closure properties of regular variation, the prior $p(\psi)$ will have regular variation.  Thus the posterior estimates of $\psi$ will also be the same as the likelihood-based estimates, provided corresponding $y_i$s are all large. For a normal prior, the prior and the likelihood have the same tail heaviness, and the results of \citet{andrade2006bayesian} imply that the posterior of $(\theta_i \mid y_i)$ will not be the same as the likelihood-based estimate. Hence, the resultant posterior estimates of $\psi$ will differ from their likelihood-based estimates, explaining Efron's observations.

From this description, it is natural to wonder if just putting a heavy-tailed prior such as a multivariate-$t$ on $\theta$ will suffice. However, in a sparse true $\theta$ case, most of the observed small $y_i$s are just noise terms. There is another key property of global-local priors that distinguishes them from other priors with heavy tails, such as a multivariate-$t$. Global-local priors also offer heavy shrinkage towards zero. Specifically, the density $p(\theta_i)$ has an unbounded spike near zero and the global shrinkage parameters adapts itself to the sparsity level of the data, when the true parameter vector is sparse. The unbounded spike near zero for horseshoe and horseshoe+ are respectively demonstrated by \citet{carvalho2010horseshoe} and in a technical report by the current authors. This is not the case with a multivariate-$t$ prior on $\theta$ and the terms with small $y_i$ are not shrunk as effectively. In presence of a true $\theta$ that is sparse-robust, that is, most components are zero and a few components are large, the heavy tails of the marginal prior on $\theta_i$ in combination with its unbounded spike near zero explains why global-local shrinkage priors perform so effectively. 


\subsection{Duality of densities} 

Since the regularly varying global-local shrinkage priors arise as normal scale mixtures, it is natural to ask what properties of the mixing distribution ensures a regularly varying right tail of the shrinkage prior, and how to generate new regularly varying shrinkage priors that are also normal scale mixtures. 

Suppose that $p(x) = \int (2\pi v)^{-1/2} e^{-x^2/2v} dF(v)$, is a normal scale mixture with moment generating function $\phi$. Theorem \ref{th:barndorff} can be used to establish the relationship between the tail behavior of the mixing distribution and the tail behavior of the normal scale mixture. 
Turning to the second question, an important tool for constructing new regularly varying priors is by using the dual densities for known normal scale mixtures. The concept of dual densities was introduced by \citet{good1995c439} as densities proportional to the characteristic functions (or moment generating functions when they exist).  Specifically, we construct
$$
\hat{p} (t) = \frac{ \phi(t)}{C}, \; \; \phi(t) = E \{\mathrm{exp}(tX)\}, \; C = \int_{\mathcal{D}} \phi(y) dy, 
$$
where $\mathcal{D}$ is such that $\hat{p}(t) \geq 0$ for all $t \in \mathcal{D}$ and $\int_{\mathcal{D}} \phi(y) dy < \infty$. 
This dual density $\hat{p}$ is also a normal scale mixture with mixing density $\hat{f}$. The mixing density for the dual and the original density denoted by $\hat{f}$ and $f$ are related by: 
$$
\hat{f}(v) = \frac{1}{(2\pi)^{1/2}p(0)v^{3/2}} f \left(\frac{1}{v}\right) , \; v > 0.
$$
The purpose of this is to be able to construct regularly varying priors that are also normal scale mixtures starting from rapidly varying priors. Dual densities were further explored by \citet{gneiting1997normal} and  \citet{nadarajah2009pdfs}. See Table \ref{tab:rv-densities} for a taxonomy of dual densities which can be used to construct slowly varying global-local scale mixtures. The classic example is the double-exponential or Laplace distribution whose dual is the Cauchy distribution. Since the Laplace can be written as a normal scale mixture with exponential mixing density, we can derive the mixing density for Cauchy.

\begin{table}[ht!]%
\centering
\def~{\hphantom{0}}
{\footnotesize
\caption{Some common exponentially tailed densities that have polynomially tailed dual densities. The densities marked with asterisks also have a known normal scale mixture representation.\label{tab:rv-densities}}
\begin{tabular}{ L{35mm}  L{45mm}  R{45mm} }
Density $p(x)$ & Dual Density $\hat{p}(x)$ & Comments \\
\hline
Exponential Power* (Special Case: Laplace, Normal) & Symmetric-stable($\alpha$) (Special case: Cauchy, Levy) & Symmetric stable distributions are heavy-tailed for $\alpha<2$. \\
\hline
Bessel function density* & Student's t & These densities are special cases of Generalized Hyperbolic distribution with parameters ($\lambda=-\frac{\nu+1}{2},\delta^2=0,\kappa^2=\nu$) and ($\lambda=-\nu/2,\delta^2=\nu,\kappa=0$) respectively. \\
\hline
Gamma (shape=$\alpha$,rate=$\lambda$) & $\hat{p}(x) = x^{-\alpha}$ & \\
\hline
Laplace* \newline Skew-Laplace I \newline Skew-Laplace II & $\hat{p}(x) \propto e^{ax}(1+x^2)^{-1}$ \newline $\hat{p}(x) \propto e^{ax}(c/(b+x)+1/x)$ \newline $\hat{p}(x) \propto \{\frac{cx}{x^2+a^2} + \frac{1}{b+x}\}$ & Heavy-tailed if $a=0.$ \newline Heavy-tailed if $a=0.$ \newline Heavy-tailed. \\
\hline
Frech\'et & $\hat{p}(x) \propto \sqrt{x}K_{-1}(a\sqrt{x})$ & Frech\'et distribution has a lower exponential tail. \\
\hline
Inverse Gaussian & $\hat{p}(x) \propto x^{1/4}K_{-\half}(a\sqrt{x})$ & \\
\hline
Linnik or $\alpha$-Laplace distribution* \newline $\alpha \in [0,2)$ & Generalized Cauchy \newline $\hat{p}(x) \propto (1+|x|^{\alpha})^{-\beta}$ & \\
\hline
\end{tabular}
}
\end{table}

\section{Examples of Nonlinear functions of multivariate normal means}\label{sec:horseshoe}


\subsection{Example 1: sum of squares} \citet{efron1973discussion} shows the following example: if $p=100, \psi= \sum_{i=1}^{100} \theta_i^2 $ and we observe $ \sum_{i=1}^{100} y_i^2 = 200 $, then the estimate $\hat{\psi} = 100 $ with standard deviation $25$ is the minimum variance unbiased estimator. It is easy to see that the posterior mean under an uninformative prior $N(0,\tau^2)$, with a large $\tau^2$, is $300$, which is badly biased upwards. This is because the marginal prior on $\psi$ heavily shrinks $\psi$ away from the origin. To be concrete, observe that
$$
\psi = \sum_{i=1}^p \theta_i^2 \sim \Gamma \left ( \frac{p}{2} , \frac{\tau^2}{2} \right ), p(\psi) = \tau^{-p} \psi^{ \frac{p}{2} -1 } e^{ - \frac{\psi}{2 \tau^2}  }.
$$
Therefore a large $\tau^2$ shrinks away from the origin due to the term $ \psi^{49}$ when $p=100$. One way of correcting this is to give $\tau$ a diffuse prior such as the half-Cauchy prior. Then the marginal prior is $p(\tau) \propto (\tau^2+1)^{-1}$ making the marginal for $\psi$ proportional to $  \psi^{-1/2} ( \psi + p )^{-1} $ and the posterior mean of $\psi$ close to the maximum likelihood estimate.  To see this, we start with the marginal prior on $\psi$, which in this case is: 
$$
p( \psi) = \int_0^\infty  \tau^{-p} \psi^{ \frac{p}{2} -1 } e^{ - \frac{\psi}{2 \tau^2}  } p(\tau^2) d \tau.
$$
As we show in Theorem \ref{th:post-ss}, the term $ p( \psi|\tau^2)$ collapses on $\tau^2 = \psi/p$ so we have a marginal prior on $\psi$ equal to the prior on $\tau^2$ evaluated at $\psi/p$. Hence we have
$$
p( \psi ) \propto \psi^{-1/2} ( \psi + p )^{-1}.
$$
This shrinks towards the origin, as the density is not zero there. This is the sense in which estimating or learning the hyperparameters helps in a non-informative analysis. \citet{berger1998estimation} points out the sub-optimality of the constant prior on $\theta$ which is also the Jeffreys' prior and derives the naive reference prior for estimating $\psi$, given by $p(\psi) = \psi^{-(p-1)/2}$. \citet{berger1999} point out the failure of profile likelihood approaches for this problem as it fails to account for the uncertainty in the nuisance parameter and show how the integrated likelihood approach avoids the pitfalls of marginalization paradox. We compare the reference prior with the other shrinkage priors and argue that the reference prior in this setting acts as a global shrinkage prior and works well in the dense case where $\theta_i = 1$ for all $i$, but its performance deteriorates in the sparse case due to over-shrinkage of large observations.


\subsection{Example 2: maximum of normal means} This example, also due to \citet{efron1973discussion}, concerns estimating $ \psi = \max \theta_i$ given observations $y_1, \cdots , y_{99}$ drawn from standard normal distribution while $y_{100}=10$.  Suppose in order to correct the problem with the sum of squares estimation under a fixed, large $\tau^2$, we have now used a prior $p(\tau^2) \propto (\tau^2 + 1)^{-2}$, as in Example 1. The posterior mean for $\psi$ is now $5$, while it is obvious that $10$ is a much more sensible estimate in this case. Now the problem arises because there is only one global shrinkage parameter $\tau$ with a heavy-tailed prior that shrinks the true signals as well as the noise terms. The James-Stein shrinkage estimator, being a purely global shrinkage estimator, suffers from the same problem. We will show that the situation can be improved by adding local shrinkage parameters in addition to a global $\tau$.

\subsection{Example 3: product of two normal means} The third marginal parameter inference problem deals with the product of normal means $\psi= \theta_1 \theta_2$ where we observe $(y_1, y_2 \mid \theta_1, \theta_2)' \sim \Nor((\theta_1,\theta_2)', \mathrm{I})$ \citep{berger1989estimating}. It follows that the reference prior in this context is proportional to $ ( \theta_1^2 + \theta_2^2 )^{-1/2}$, which has polynomially decaying tails like the half-Cauchy prior, but is not integrable. For a detailed review of reference prior for product normal means and its generalization to the multivariate case, we refer the readers to \citet{sun1995reference, sun1999reference}. We will show when both $\theta_1$ and $\theta_2$ are close to zero, the posterior distribution under the global-local priors concentrates around the origin, unlike its competitors, including normal, pure global and pure local shrinkage priors. 

\subsection{Example 4: ratio of two normal means}  Finally, we consider the Fieller--Creasy problem, where the parameter of interest is $\psi= \theta_1/\theta_2$ \citep{fieller1940biological,creasy1954limits}.  Defining $\lambda = \theta_2$ as the nuisance parameter, the reference prior is obtained as $ p(\psi, \lambda) \propto (\psi^2 +1 )^{-1/2} $ \citep{liseo1993elimination} which is Cauchy-like but not integrable. One expects horseshoe to produce similar results and as we show in \S \ref{sec:pm-fc}, these priors lead to sensible answers by concentrating around the true value for $\theta_2 \neq 0$ and around the origin when $\theta_1 = \theta_2 = 0$. In all the examples, global-local priors enjoy the benefit of being integrable.

\section{Results}\label{sec:results}
\subsection{Simulation scheme} We compare the following priors for the marginal parameter of interest problems of \S \ref{sec:horseshoe}.
\ben
\item The global-local shrinkage priors, namely, the horseshoe and the horseshoe+ priors. %
\item The Laplace or double-exponential prior, with the mixing density given by: 
\begin{align}
p(\lambda_i^2 \mid \tau^2) & = (2\tau^2)^{-1}{\exp}\{ - \lambda_i^2/2\tau^2 \}, \label{eq:laplace-1} \\
 (\tau^2 \mid \xi, d) & \sim {\mathrm IG}(\xi/2, \xi d^2/2), \xi >0, d > 0. \label{eq:laplace-2}
\end{align}
The Laplace prior has a long history in Bayesian robustness modeling \citep{pericchi1992exact, box1962further}. For the numerical examples, we use the standard double-exponential prior, that is, fix $\xi = d = 1$.
\item The vague normal prior, that is, $\theta_i \sim \Nor(0, \sigma^2 = 300)$. 
\item The pure-local shrinkage prior, and the pure-global shrinkage priors, by taking $\tau  = 1$ or, $\lambda_i = 1$, for all $i = 1, \ldots, p$, in the hierarchical model for horseshoe prior in \S 1.2. These two priors are included to highlight the role of both local and global shrinkage parameters. 
\item The reference priors for Examples 1, 3 and 4. 
\een
In the following simulations, the estimates are produced using the probabilistic programming language \texttt{Stan} \citep{stan-software:2014} according to the parametrization shown above. \texttt{Stan} is run with 4 parallel chains of 10,000 iterations. We also provide two efficient algorithms for sampling from the horseshoe for horseshoe+ posteriors in the Supplementary Material. 

\subsection{Sum of squares estimation}\label{sec:sum}
\begin{figure}[ht!]
\centering
\begin{subfigure}[b]{0.45\textwidth}%
\raggedleft
\includegraphics[width=\linewidth]{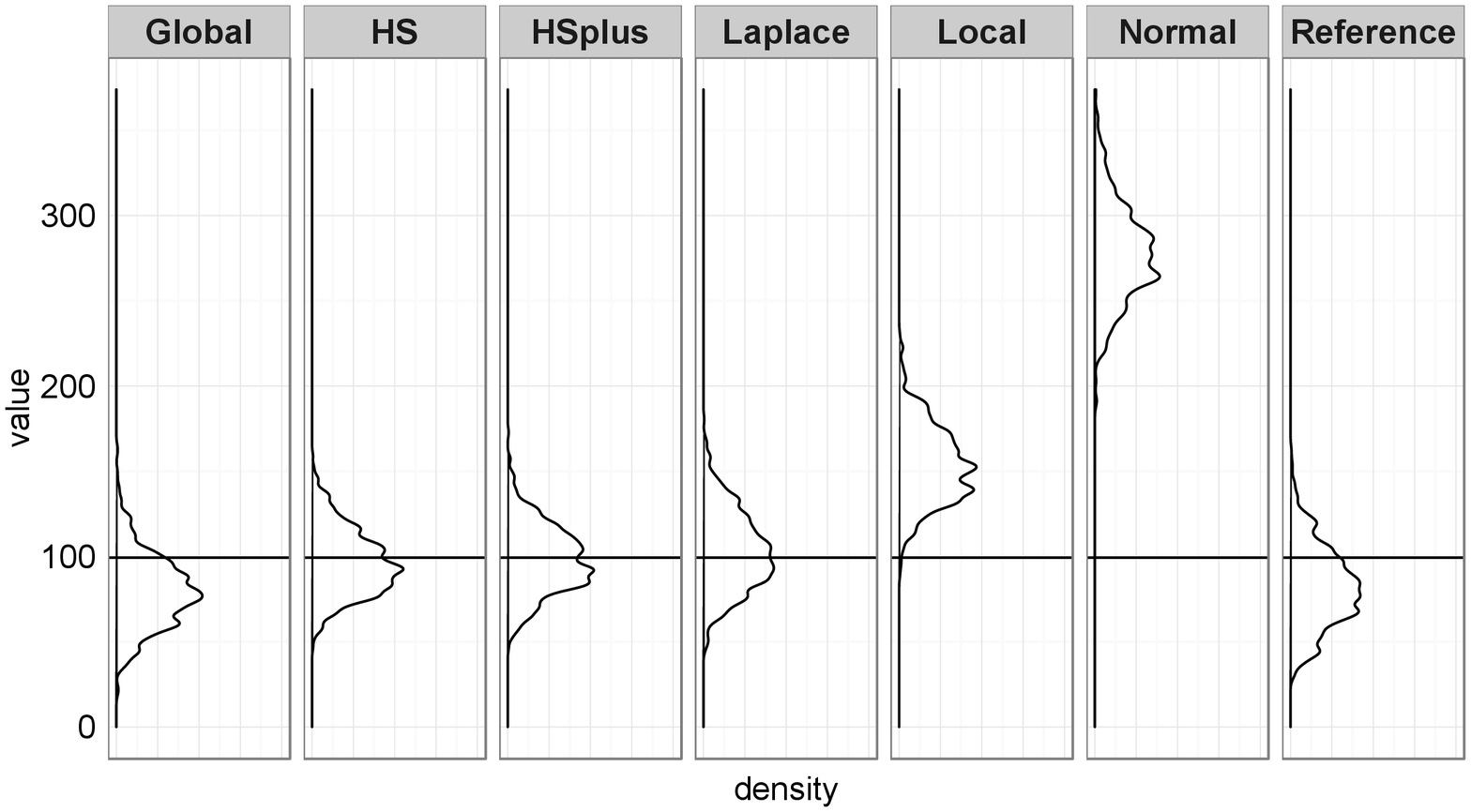}%
\caption{$A = 10$ and $q_p = 1$.}%
\label{fig:efron-ss-sparse-1}%
\end{subfigure}
\begin{subfigure}[b]{0.45\textwidth}%
\raggedright
\includegraphics[width=\linewidth]{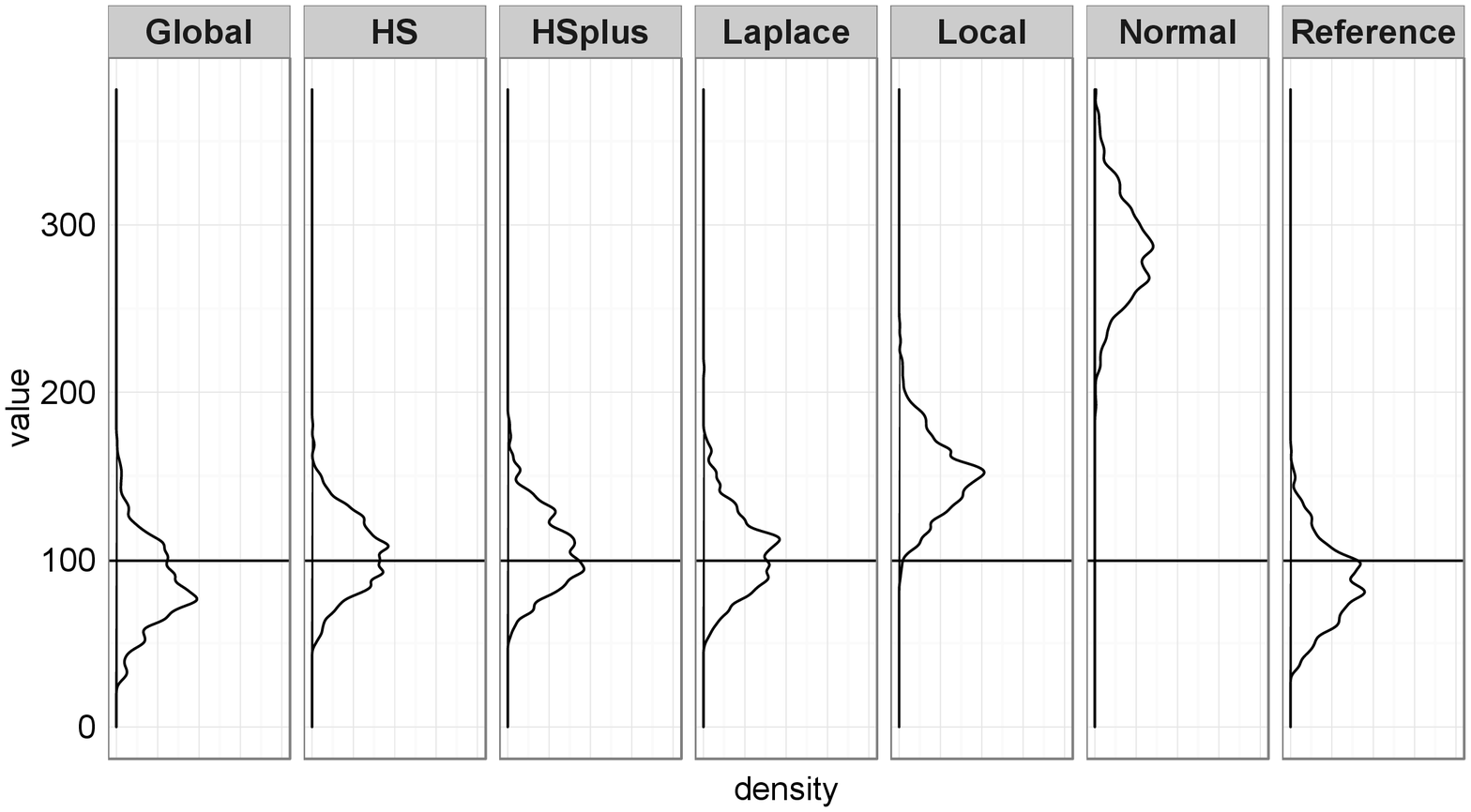}%
\caption{$A = 5$ and $q_p = 4$ .}%
\label{fig:efron-ss-sparse-2}%
\end{subfigure}
\begin{subfigure}[b]{0.45\textwidth}%
\centering
\includegraphics[width=\linewidth]{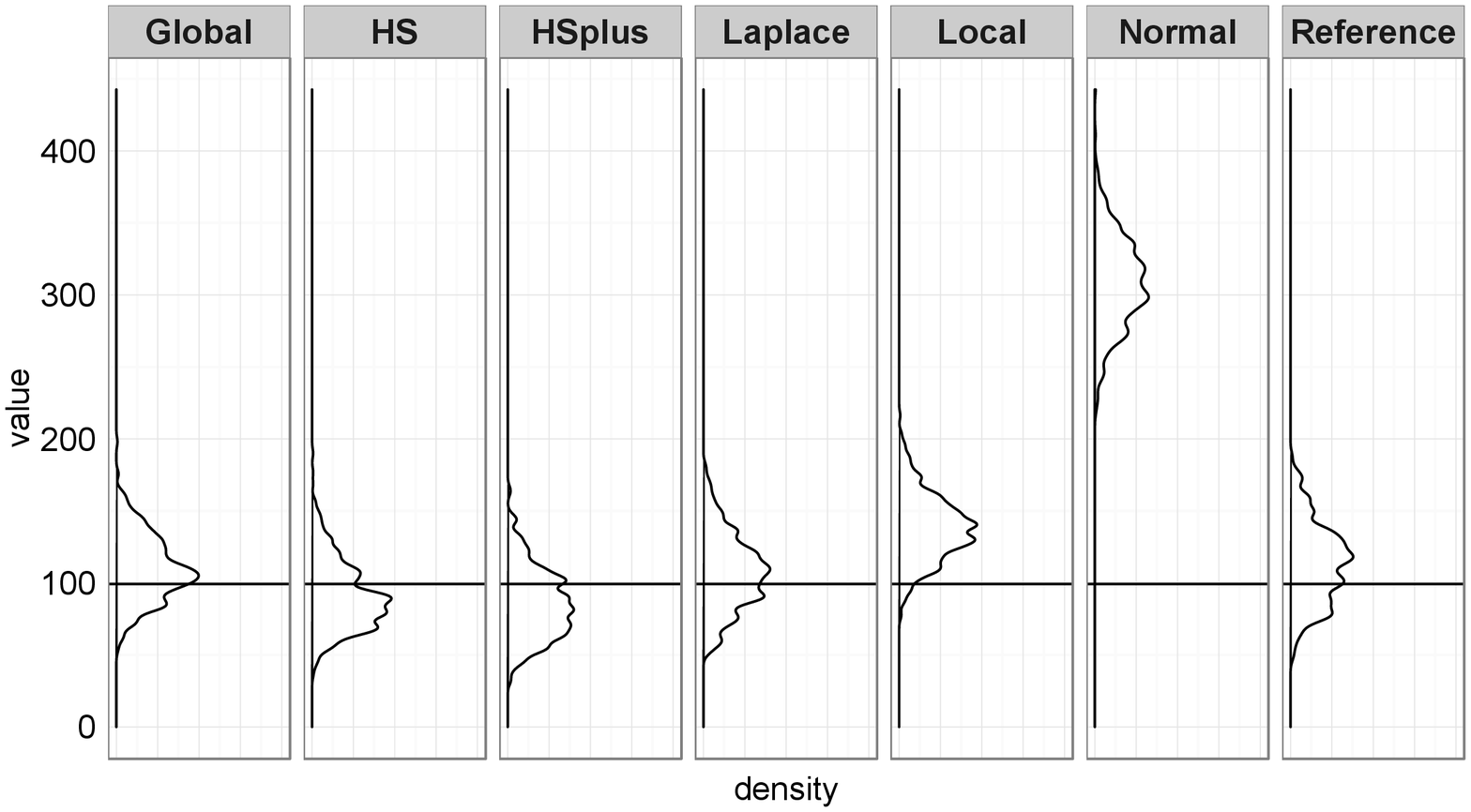}%
\caption{$A = 1$ and $q_p = 100$.}%
\label{fig:efron-ss-sparse-3}%
\end{subfigure}
\caption{Posterior densities of the parameter of interest $\psi = \sum_{i=1}^{100} \theta_i^2$, for the candidate priors, namely, the horseshoe+ (HSplus), the horseshoe (HS), Laplace, normal, pure-local, pure-global and the reference priors. In each panel, $q_p$ is the number of non-zero means and $A$ is the magnitude of the non-zero means. The horizontal line is at the true value $\psi = 100$. }
\label{fig:efron-ss}
\end{figure}

For comparing the candidate priors, we chose $\theta$ with $q_p$ non-zero components of magnitude $A$ and dimension $p = 100$. The simulation study shows the performance for  different $\theta$ in both the sparse regime where $q_p/p \to 0$ and completely dense where $q_p/p \to 1$. Table \ref{tab:efron-ss} shows the performance comparison for the candidate priors for three different sparsity patterns while constraining $\theta$ to have unity quadratic mean, that is, $\vectornorm{\theta}^2 = p$. The chosen values for $A$ and $q_p$ are $A = 10, q_p = 1$, $A = 5, q_p = 4$, and $A = 1, q_p = 100$ respectively. The posterior quantiles, mean and standard deviation for $\psi = \sum_{i=1}^{p} \theta_i^2$ for the three experiments are given in Table \ref{tab:efron-ss}, and Figure \ref{fig:efron-ss} shows the corresponding posterior distribution of $\psi$. 

The first two blocks of Table \ref{tab:efron-ss} and the sub-figures Fig. \ref{fig:efron-ss-sparse-1} and Fig. \ref{fig:efron-ss-sparse-2} shows the relative performance of the different priors in the two sparse situations, that is, $A = 10, q_p = 1$, and $A = 5, q_p = 4$. As expected, both horseshoe and horseshoe+ gives sensible answers in the sparse case with the posterior distribution centered near $100$ and standard deviation close to $25$ as desired by \citet{efron1973discussion}. The Laplace prior also does well for the sparse $\theta$, but as we shall see shortly, its performance deteriorates for the remaining examples, maximum, product of means and the Fieller--Creasy problem, advocating against its use as a default prior. The density plots Fig. \ref{fig:efron-ss-sparse-1} and Fig. \ref{fig:efron-ss-sparse-2} are instructive to see where the badly performing priors concentrate. The vague normal prior leads to an upward bias in all cases, as \cite{efron1973discussion} had pointed out. Interestingly, the pure-local shrinkage prior concentrates above as it can not adapt to the sparsity in absence of a global shrinkage parameter. On the other hand, pure-global prior and the reference prior shrinks the posterior more towards zero and lead to under-estimation, as one should expect when the only shrinkage mechanism is a global parameter. 

The dense case is $\theta_i = 1$, for all $ i = 1, \ldots, p$. The global-local shrinkage priors, that clearly outperform the others in the sparse case, are not the best performers here as they are designed to shrink the small $\theta_i$s, which under-estimates $\sum_{i=1}^{p} \theta_i^2$. The pure-global shrinkage prior is the best performer in the dense case, followed closely by the reference prior. The global prior leads to a posterior that concentrates on $(0, \sum_{i=1}^{p} y_i^2)$, unlike the vague normal prior that concentrates above this region. We prove this theoretically in \S \ref{sec:post-conc}. 

To summarize, the global-local shrinkage priors give a more sensible answer than the proper flat normal prior with a high variance in the sparse case. The normal prior not only gives a posterior mean far away from the true value, as Fig. \ref{fig:efron-ss} shows, the posterior for a vague normal prior puts most of its density on a different region, unlike the global-local shrinkage priors that concentrates near the true $\psi$. Even in the dense case, where the global-local shrinkage priors shrink too strongly, a global shrinkage prior with a half-Cauchy prior produces an accurate estimates, but one cannot trade the half-Cauchy with a $\Nor(0, 300)$ prior, suggesting regular variation as a key property for Bayesian robustness. 


\begin{table}[ht!]
  \centering
  \caption{Summary statistics and standard deviation of the posterior distribution for the candidate priors, namely, the horseshoe+ (HS+), the horseshoe (HS), Laplace, normal, pure-local, pure-global and the reference priors, for parameter of interest $\psi = \sum_{i=1}^{100} \theta_i^2$. Here $q_p$ is the number of non-zero means and $A$ is the magnitude of the non-zero means. True $\psi = 100$ in all cases.}
	{\footnotesize
\begin{tabular}{rrrrrrrrr}
       &       & Minimum  & $Q_1$ & Median  & Mean  & $Q_3$ & Max   & Std. Dev.  \\
A = 10, $q_p = 1$ & HS+   & 46${\cdot}$7  & 84${\cdot}$4  & {97${\cdot}$6} & {98${\cdot}$3} & 111${\cdot}$6 & 172${\cdot}$8 & 19${\cdot}$8 \\
      & HS    & 46${\cdot}$6  & 82${\cdot}$3  & {94${\cdot}$6} & {96${\cdot}$6} & 108${\cdot}$7 & 159${\cdot}$0 & 19${\cdot}$2 \\
      & Laplace & 44${\cdot}$9  & 86${\cdot}$7  & {101${\cdot}$7} & {103${\cdot}$5} & 119${\cdot}$3 & 180${\cdot}$4 & 23${\cdot}$2 \\
      & Normal & 189${\cdot}$7 & 260${\cdot}$0 & 276${\cdot}$9 & 278${\cdot}$0 & 295${\cdot}$3 & 373${\cdot}$9 & 29${\cdot}$4 \\
      & Local & 89${\cdot}$5  & 137${\cdot}$8 & 152${\cdot}$4 & 153${\cdot}$9 & 169${\cdot}$2 & 229${\cdot}$7 & 22${\cdot}$7 \\
      & Global & 19${\cdot}$5  & 64${\cdot}$6  & 78${\cdot}$9  & 79${\cdot}$9  & 93${\cdot}$1  & 164${\cdot}$1 & 21${\cdot}$6 \\
      & Reference & 29${\cdot}$4  & 66${\cdot}$3  & 81${\cdot}$1  & 82${\cdot}$5  & 98${\cdot}$0  & 162${\cdot}$8 & 23${\cdot}$9 \\
      &       &       &       &       &       &       &       &  \\
      &       & Minimum  & $Q_1$ & Median  & Mean  & $Q_3$ & Max   & Std${\cdot}$ Dev${\cdot}$  \\
A = 5, $q_p = 4$ & HS+   & 53${\cdot}$2  & 89${\cdot}$8  & {104${\cdot}$0} & {105${\cdot}$9} & 121${\cdot}$4 & 182${\cdot}$8 & 22${\cdot}$8 \\
      & HS    & 49${\cdot}$8  & 88${\cdot}$2  & {103${\cdot}$1} & {103${\cdot}$6} & 119${\cdot}$0 & 180${\cdot}$3 & 21${\cdot}$6 \\
      & Laplace & 50${\cdot}$6  & 87${\cdot}$9  & {104${\cdot}$0} & {105${\cdot}$0} & 119${\cdot}$2 & 214${\cdot}$2 & 23${\cdot}$8 \\
      & Normal & 191${\cdot}$6 & 265${\cdot}$5 & 285${\cdot}$3 & 285${\cdot}$7 & 305${\cdot}$2 & 380${\cdot}$9 & 30${\cdot}$8 \\
      & Local & 88${\cdot}$9  & 134${\cdot}$4 & 149${\cdot}$9 & 150${\cdot}$3 & 164${\cdot}$7 & 240${\cdot}$0 & 23${\cdot}$5 \\
      & Global & 27${\cdot}$6  & 70${\cdot}$1  & 84${\cdot}$0  & 86${\cdot}$2  & 102${\cdot}$5 & 170${\cdot}$6 & 24${\cdot}$8 \\
      & Reference & 34${\cdot}$36 & 69${\cdot}$64 & 85${\cdot}$3  & 86${\cdot}$5  & 100${\cdot}$9 & 164${\cdot}$7 & 23${\cdot}$7 \\
      &       &       &       &       &       &       &       &  \\
      &       & Minimum  & $Q_1$ & Median  & Mean  & $Q_3$ & Max   & Std${\cdot}$ Dev${\cdot}$  \\
A = 1 , $q_p = 100$ & HS+   & 31${\cdot}$9  & 67${\cdot}$8  & 83${\cdot}$8  & 86${\cdot}$0  & 101${\cdot}$8 & 165${\cdot}$6 & 24${\cdot}$1 \\
      & HS    & 36${\cdot}$3  & 73${\cdot}$8  & {88${\cdot}$5} & {91${\cdot}$6} & 107${\cdot}$2 & 190${\cdot}$0 & 24${\cdot}$0 \\
      & Laplace & 51${\cdot}$2  & 89${\cdot}$6  & {107${\cdot}$1} & {107${\cdot}$6} & 123${\cdot}$7 & 181${\cdot}$9 & 26${\cdot}$3 \\
      & Normal & 219${\cdot}$3 & 288${\cdot}$0 & 308${\cdot}$8 & 309${\cdot}$0 & 331${\cdot}$4 & 442${\cdot}$6 & 32${\cdot}$5 \\
      & Local & 76${\cdot}$4  & 123${\cdot}$7 & 138${\cdot}$0 & 138${\cdot}$4 & 153${\cdot}$2 & 216${\cdot}$3 & 23${\cdot}$3 \\
      & Global & 53${\cdot}$23 & 94${\cdot}$73 & {108${\cdot}$0} & {110${\cdot}$2} & 125${\cdot}$7 & 198   & 22${\cdot}$9 \\
      & Reference & 47${\cdot}$67 & 92${\cdot}$45 & 114${\cdot}$5 & 113${\cdot}$8 & 131${\cdot}$5 & 188${\cdot}$3 & 27${\cdot}$8 \\
\end{tabular}%
		}
  \label{tab:efron-ss}%
\end{table}%

%
%


\subsection{Maximum estimation}\label{sec:max}

\begin{figure}[ht!]%
\centering
\includegraphics[width=0.5\linewidth,height = 2in]{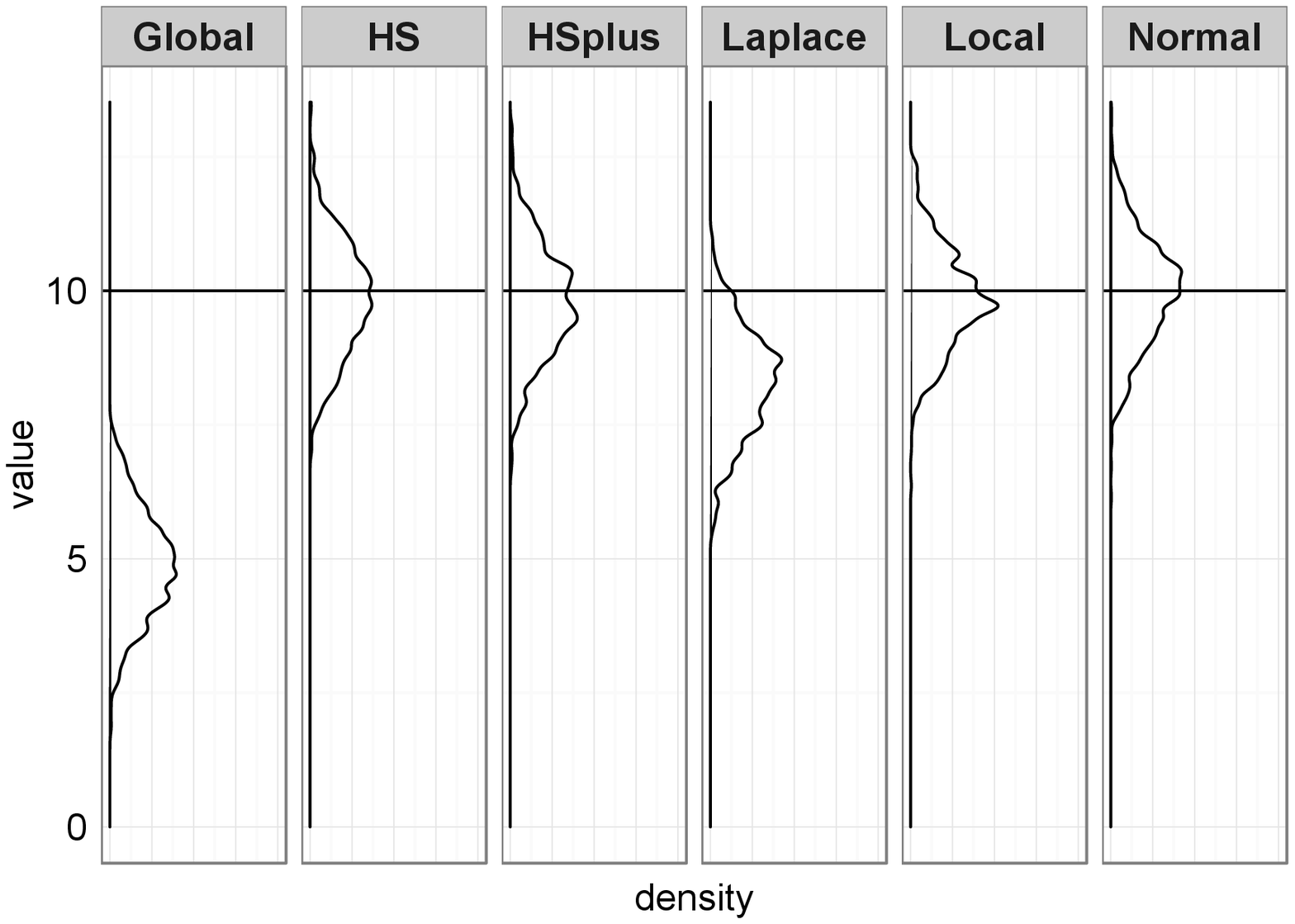}%
\caption{Posterior densities of the parameter of interest $\psi = \max \theta_i$, for the candidate priors, namely, the horseshoe+ (HSplus), the horseshoe (HS), Laplace, normal, pure-local and pure-global priors, for $(y_i \mid \theta_i) \sim \Nor(0,1)$, $i = 1, \ldots, 99$ and $y_{100} = 10$. The horizontal line is at $y_{\max}$.}%
\label{fig:efron-max}%
\end{figure}

Here we show that the horseshoe priors do well for estimating $\theta_{(p)}$, the maximum element of $\theta$, when one $y_i$ is an outlier at $10$ and the rest of the $y_i$s are standard normal variates. We expect that the global-local shrinkage priors to do well due to the local shrinkage parameters that provide robustness in the tails, that is $E(\theta \mid y) \approx y$ for large $|y|$, whereas a pure-global shrinkage prior results in a posterior mean of around $5$, as pointed out by \citet{efron1973discussion}. Indeed, as Table \ref{tab:efron-max} and Fig. \ref{fig:efron-max} show, both the global-local priors as well as the normal prior with high standard deviation do well for this example while the worst-performing candidate is the global shrinkage prior. Interestingly, the Laplace prior performs poorly due to its over-shrinkage on the tails. This agrees with fact that the Laplace prior leads to a biased estimate for large observations, as noted in the discussion in \S 3.2 of \cite{carvalho2010horseshoe}, where they showed that for large $|y_i|$ there is a constant bias, $|E(\theta_i | y_i) - y_i | \approx \sqrt{2}/\tau$, for a given $\tau$. 

\begin{table}[ht!]
  \centering
  \caption{Summary statistics and standard deviation of the posterior distribution for parameter of interest $\psi = \max \theta_i$, for the candidate priors, namely, the horseshoe+ (HS+), the horseshoe (HS), Laplace, normal, pure-local and pure-global priors, for $[y_i \mid \theta_i] \sim \Nor(0,1)$ for $i = 1, \ldots, 99$ and $y_{100} = 10$.}
	{\footnotesize 
\begin{tabular}{rrrrrrrrr}
      &       & Minimum  & $Q_1$ & Median  & Mean  & $Q_3$ & Max   & Std.Dev.  \\
 & HS+   & 6$\cdot$7   & 9$\cdot$0   & 9$\cdot$7   & 9$\cdot$7   & 10$\cdot$4  & 12$\cdot$5  & 1$\cdot$1 \\
      & HS    & 6$\cdot$8   & 9$\cdot$2   & 9$\cdot$8   & 9$\cdot$9   & 10$\cdot$5  & 12$\cdot$9  & 1$\cdot$0 \\
      & Laplace & 4$\cdot$8   & 7$\cdot$8   & 8$\cdot$5   & 8$\cdot$4   & 9$\cdot$1   & 11$\cdot$0  & 1$\cdot$0 \\
      & Normal & 6$\cdot$6   & 9$\cdot$3   & 10$\cdot$0  & 9$\cdot$9   & 10$\cdot$6  & 13$\cdot$1  & 1$\cdot$0 \\
      & Local & 6$\cdot$7   & 9$\cdot$1   & 9$\cdot$9   & 9$\cdot$8   & 10$\cdot$5  & 13$\cdot$0  & 1$\cdot$0 \\
      & Global & 1$\cdot$5   & 3$\cdot$7   & 4$\cdot$4   & 4$\cdot$4   & 5$\cdot$1   & 7$\cdot$7   & 1$\cdot$0 \\
\end{tabular}%
}
\label{tab:efron-max}%
\end{table}%

\subsection{Bivariate normal means: product and ratio}\label{sec:pm-fc}

\begin{figure}[ht!]%
\centering
\includegraphics[width=0.8\linewidth,height=2.5in]{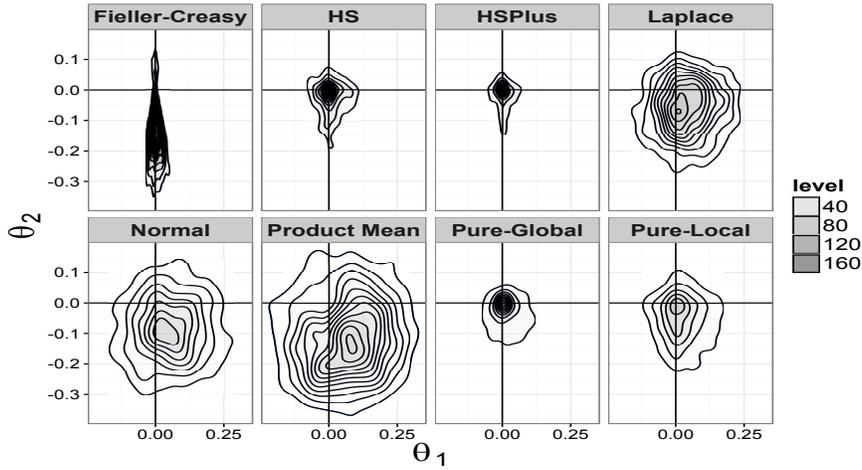}%
\caption{Two-dimensional contour plots of $p(\theta_1, \theta_2 \mid y)$ for the candidate priors, namely, the horseshoe+ (HSPlus), the horseshoe (HS), Laplace, normal, pure-local, pure-global and the two reference priors for the product mean and ratio of two means (Fieller-Creasy) problems. }
\label{fig:fc_plot_theta_density}%
\end{figure}

We compare the different shrinkage priors for Examples 3 and 4 in \S \ref{sec:horseshoe}, concerning the product and ratio of means \citep{berger1989estimating, creasy1954limits} of a bivariate normal distribution. The data are generated as $K = 100$ independent draws from a bivariate normal distribution as
\begin{align*}
  (y_k \mid \theta_1, \theta_2) &\sim \Norm \left((\theta_1, \theta_2), \mathrm{I} \right), k = 1, \ldots, K. 
\end{align*}
The sample mean for the data is $(\bar{y}_1, \bar{y}_2) = (0{\cdot}0427, -0{\cdot}0840)$. The results below pertain to the case where the true values for both $\theta_1$ and $\theta_2$ are zero. 
Figure~\ref{fig:fc_plot_theta_density} shows the two-dimensional density contours for all the candidate posteriors. Fig.~\ref{fig:fc_plot_theta_density} clearly illustrates that the posterior distributions under the horseshoe and the horseshoe+ priors are highly concentrated in a small interval near the origin unlike their competitors, such as the double-exponential, the normal, or the pure-local and pure-global shrinkage priors. As a result, the global-local shrinkage priors enjoy higher estimation accuracy compared to the competitors. The global shrinkage estimator also concentrates near the origin but with a higher spread, implying regular-variation is a key property for learning marginal parameters of interest. 

\begin{table}[ht!]
\centering
\def~{\hphantom{0}}
\caption{Posterior quantiles for the marginal parameters of interest $\psi_1 = \theta_1/\theta_2$ and $\psi_2 = \theta_1 \theta_2$ for the candidate priors, namely, the horseshoe+ (HS+), the horseshoe (HS), Laplace, normal, pure-local, pure-global and the reference priors. All entries in the table are multiplied by $1,000$.}{%
{\footnotesize
\begin{tabular}{rrrrrrrr}
      & \multicolumn{3}{c}{$\psi_1 = \theta_1/\theta_2$} &       & \multicolumn{3}{c}{$\psi_2 = \theta_1 \theta_2 $} \\
      & $Q_1$ & Median  & $Q_3$ &       & $Q_1$ & Median  & $Q_3$ \\
HS+   & -1020 & -24${\cdot}$8 & 696${\cdot}$7 & HS+   & -0${\cdot}$6  & -0${\cdot}$0  & 0${\cdot}$2 \\
HS    & -941${\cdot}$3 & -37${\cdot}$8 & 689${\cdot}$7 & HS    & -1${\cdot}$1  & -0${\cdot}$0  & 0${\cdot}$5 \\
Laplace & -1092 & -170${\cdot}$2 & 618${\cdot}$1 & Laplace & -5${\cdot}$2  & -0${\cdot}$4  & 2${\cdot}$1 \\
Normal & -1060 & -250${\cdot}$3 & 482${\cdot}$2 & Normal & -8${\cdot}$6  & -1${\cdot}$3  & 2${\cdot}$4 \\
Local & -999${\cdot}$6 & -126${\cdot}$7 & 489${\cdot}$2 & Local & -3${\cdot}$9  & -0${\cdot}$2  & 1${\cdot}$3 \\
Global & -1092 & -120${\cdot}$5 & 812${\cdot}$1 & Global & -2${\cdot}$1  & -0${\cdot}$0  & 0${\cdot}$8 \\
Reference & -71${\cdot}$9 & -8${\cdot}$5  & 58${\cdot}$7  & Reference & -14${\cdot}$6 & -3${\cdot}$8  & 3${\cdot}$9 \\
\end{tabular}%
}%
}
\label{tab:fc-and-pm}
\end{table}
\section{Posterior contraction for the sum of squares problem}\label{sec:post-conc}
Here we provide some theoretical basis to the behavior of horseshoe prior for the sum of squares problem. It turns out that a sufficient statistic for $\psi = \sum_{i=1}^{p} \theta_i^2$ is $Z = \sum_{i=1}^{p} y_i^2$ and $Z$ follows a non-central $\chi^2$ distribution with $p$ degrees of freedom and non-centrality parameter $\psi$ with density:
\begin{equation*}
f_Z(z \mid \psi) = \frac{1}{2} e^{-(z+\psi)/2}\left(\frac{z}{\psi} \right)^{p/4-1/2} I_{p/2-1}(\sqrt{\psi z}),
\end{equation*}
where $I_{\nu}(z)$ is the modified Bessel function of the first kind. This allows us to investigate the properties of the posterior under two different priors on the non-centrality parameter $\psi$, namely, the prior induced on $\psi$ by putting a horseshoe prior and a normal prior on each parameter $\theta_i$. The limiting form $I_{\nu}(z) \sim e^z/\sqrt{2\pi z}$ when $\nu$ is fixed and $z \to \infty$ \citep[Chapter 10.30 (ii) of][]{Olver:2010:NHMF} implies
\begin{equation}
f_Z(z \mid \psi) \approx \frac{1}{2\sqrt{2 \pi}} e^{-(\sqrt{z}-\sqrt{\psi})^2/2}z^{(p-3)/4}\psi^{-(p-1)/4},\label{eq:lik}
\end{equation}
for large $z$. We can use this approximation to prove the following:
\begin{theorem}\label{th:post-ss}
The posteriors for $\psi$ as $Z\to \infty$ under the horseshoe prior and the gamma prior induced by normal prior on $\theta_i$s are approximately given by
\begin{align}
p_{HS}(\psi \mid Z) & \propto e^{-(\sqrt{z}-\sqrt{\psi})^2/2}\psi^{-(p+1)/4} \frac{1}{(\psi+p)}, \label{eq:hs_ss_post} \\
p_{N}(\psi \mid Z) & \propto e^{-(\sqrt{z}-\sqrt{\psi})^2/2}\psi^{(p-3)/4} e^{ - \frac{\psi}{2 \tau^2}  }. \label{eq:psi-norm}
\end{align}
\end{theorem}

\begin{figure}[h!]
\centering 
\includegraphics[width=0.7\linewidth,height=2in]{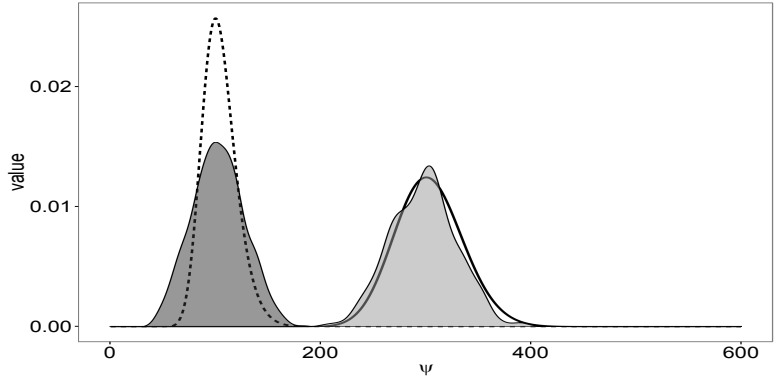}
\caption{Kernel density estimates (shaded regions) of the posterior $p(\psi \mid Z)$ under the half-Cauchy and the $\Nor(0,300)$ prior on each $\theta_i$ along with their Laplace approximations (lines) given by \eqref{eq:hs_ss_post} and \eqref{eq:psi-norm}. The darker shade and dotted line denote the $\Nor(0,300)$ prior and the lighter shade and solid line denote the half-Cauchy prior. }
\label{fig:comp.psi}
\end{figure}
The main difference between the horseshoe prior and the gamma prior is that the posterior density under the horseshoe prior is non-zero at $\psi=0$, unlike the posterior density under the gamma prior. In fact, the posterior density under the horseshoe prior has a pole at $\psi=0$ that shrinks the posterior mean $E(\psi \mid Z)$ towards the origin. The kernel density estimates of both the posterior densities are shown in Fig. \ref{fig:comp.psi}, along with their approximations derived in \eqref{eq:hs_ss_post} and \eqref{eq:psi-norm}. The plot serves two purposes. First, it shows the accuracy of the Laplace approximation. Secondly, it shows that the posterior density under horseshoe prior and the gamma prior concentrates near $\psi = 100$ and $\psi = 300$, and the two posterior distributions concentrates on two disjoint regions, namely, horseshoe on $(0,Z)$ and gamma on $(Z,\infty)$. In fact, this can be proved theoretically. We shall now show that the posterior distribution of $\psi$ for horseshoe prior is going to put almost all of its density in the interval $(0,Z)$, and the posterior for gamma prior is contained almost entirely in the region $[Z, \infty)$ for large values of $\tau^2$.

\begin{theorem}\label{th:horse-ss}
Let $[y_i \mid \theta_i] \sim \Nor(\theta_i,1)$ and $\psi = \sum_{i=1}^{p} \theta_i^2$,  and $Z = \sum_{i=1}^{p} y_i^2$. Let the prior on $\theta_i$ be the horseshoe prior given by the hierarchy: $\theta_i \sim \Nor(0,\tau^2)$ and $\tau \sim C^+(0,1)$. Then, for large $Z$ the posterior distribution of $\psi$ given $Z$ is concentrated on the interval $(0,Z)$, that is,
$$
P(\psi \geq Z \mid Z ) \leq 4(p+1)^{-1} \mbox{ as } Z \to \infty.
$$
\end{theorem}
Now we show that the posterior mean under the gamma prior induced by a component-wise $\Nor(0,\tau^2)$ converges to $\sum_{i=1}^{p} y_i^2 + p$ as $\tau^2 \to \infty$. In fact, we prove a stronger result that $p(\psi \mid Z)$ shrinks away from the interval $(0,Z)$ as $p \to \infty$ if $\tau^2$ is large enough such that $\sum_{i=1}^{p} y_i^2  < 2\tau^2$.
\begin{theorem}\label{th:norm-ss}
Let $[y_i \mid \theta_i] \sim \Nor(\theta_i,1)$ and $\psi = \sum_{i=1}^{p} \theta_i^2$,  and $Z = \sum_{i=1}^{p} y_i^2$. Let the prior on $\theta_i$ be the normal prior: $\theta_i \sim \Nor(0,\tau^2)$ where $\tau^2$ is large such that $Z \leq 2\tau^2$. Then, for large $p$ the posterior distribution of $\psi$ given $Z$ is concentrated on the interval $(Z,\infty)$, that is, 
$$
P(\psi \leq Z \mid Z ) \leq \frac{1}{\Gamma(\frac{p+5}{4})} \left(\frac{Z}{2\tau^2} \right)^{\frac{p+1}{4}} \mbox{ as } p \to \infty.
$$
\end{theorem}
The proofs of the theorems are given in the Supplementary Material. We conjecture similar concentration results will hold for the horseshoe+ prior, although an investigation is beyond the scope of the current article.

\section{Discussion}\label{sec:discuss}
Global-local shrinkage priors such as the horseshoe and horseshoe+ priors were initially designed as a computationally attractive alternative to spike-and-slab priors for the sparse signal recovery problem, where pure-global shrinkage estimators such as the James-Stein estimator perform poorly \citep{polson2010shrink}. Several optimality results are available for the estimation of normal means using these priors \cite[for example,][]{van2014horseshoe,ghosh2015asymptotic}. The goal of the current article is to demonstrate that these priors go beyond estimating just the normal means; they also perform well in estimating low-dimensional, nonlinear functions of interest. We show numerically that the class of regularly varying priors performs as well as Bernardo's reference prior in the bivariate product and ratio problems, and better for the sum of squares problem in a sparse setting, and thus satisfies the informal criterion for a candidate default prior ``which never leads to an obviously stupid answer'' \citep{scott2014comment}. On one hand, the reference priors are quite general in that they formally maximize the divergence between the prior and the posterior, and it is not clear whether the regular varying global-local priors possess any such properties. On the other hand, (a) the computing reference prior can be very cumbersome for many different problems (and may not even be possible), (b) if one changes the parameterization reference prior needs to be recomputed, (c)  reference priors are often not integrable, making them unsuitable for model selection purposes  and (d) reference priors depend on an ordering of the model parameters, even in cases where no such natural ordering exists.  A formal non-informative prior is not always appropriate or useful, for example when the true $\theta$ is sparse \citep{scott2006exploration}, where the global-local shrinkage prior is optimal in terms of Bayes risk and asymptotic minimaxity \citep{van2014horseshoe,ghosh2015asymptotic}. The global-local priors also come with the added benefit of always being proper. Thus, it can be intuitively argued that the regularly varying global-local priors offer a practical compromise between Bernardo's formal framework and flat priors. It should be pointed out that although regular variation, along with global-local shrinkage, appears to be a key property for success in multi-parameter problems, a thorough formal investigation should be considered future work.

\section*{Acknowledgment}
The authors thank the Editor, the Associate Editor and two anonymous referees for their constructive suggestions. The second author was supported by the U.S. National Science Foundation through the Statistical and Applied Mathematical Sciences Institute. 

\beginsupplement{
\section*{Supplementary material}
\label{SM}
The supplementary file contains more details on regular variation, the proofs of theorems in \S \ref{sec:post-conc}, sampling strategies for the global-local shrinkage priors, additional simulation results and a discussion of connections with the penalized complexity priors. 
\section{Regular variation}\label{sec:regvar}
The concept of regular variation is one of the most used tools in applied probability and is discussed in great details in Chapter VIII of \citet{feller1971} and by \citet{bingham1989regular}. 
\begin{Def}
A measurable function $f(\cdot)$ is regularly varying at $\infty$ if with index $\alpha \in \Re$, denoted as $f \in R_{\alpha}$, if 
\beq
\lim_{x \to \infty} \frac{f(\lambda x)}{f(x)} \to \lambda^\alpha, \quad \forall \lambda > 0. \label{eq:regvar}
\eeq
Similarly, a measurable function $f(\cdot): \Re^{+}\to \Re^{+} $ is regularly varying at the origin if with index $\alpha \in \Re$, denoted as $f \in R_{\alpha}(0+)$, if 
\beq
\lim_{x \to 0+} \frac{f(\lambda x)}{f(x)} \to \lambda^\alpha, \quad \forall \lambda > 0. \label{eq:regvar0}
\eeq
\end{Def}
Using the characterization theorem in \citet{bingham1989regular}, it can be shown that, $f \in R_{\alpha}$ if and only if $f(x) = x^{\alpha}l(x)$, where $l \in R_0$, and such a function $l(\cdot)$ is called a slowly-varying function. One can analogously define rapid variation where a measurable function $f$ is rapidly varying with index $\infty$ when the limit of the quantity on the left hand side of Equation (\ref{eq:regvar}) goes to infinity. 

\begin{Def}
A random variable $Z$ has a regularly varying right tail if with index $\alpha \in \Re$ if its distribution (say $F_Z$) satisfies: 
\begin{equation*}
1-F_Z(x) = P(Z > x) = x^{-\alpha} l(x), \quad \mbox{for } x > 0.
\end{equation*}
In other words, $Z \sim F_Z$ is $RV(\alpha)$ if $\bar{F}_Z \in R_{-\alpha}$. 
\end{Def}

One of the most important properties of regular variation is that it is preserved under many operators that commonly arise in inferential problems where we are interested in a many-to-one transformation of a parameter vector. A few of the closure properties that appear in Proposition 1.5.7 of \citet{bingham1989regular} are listed below.

\begin{proposition}\label{lemma:closure}
\begin{enumerate}
	\item if $f(x) \in R_{\alpha}$, then $f(x)^\rho \in R_{\rho \alpha}$. 
	\item if $f_i \in R_{\alpha_i}$, for $i = 1,2$, then $f_1+ f_2 \in R_{\alpha}$ where $\alpha = \max\{\alpha_1, \alpha_2\}$. 
	\item $f_i \in R_{\alpha_i}$, for $i = 1,\ldots,k$, and if $r(x_1, \ldots, x_k)$ is a rational function with positive coefficients then $r(f_1(x), \ldots, f_k(x)) \in R_{\alpha}$ for some $\alpha$. 
\end{enumerate}
\end{proposition}
Similar closure properties can be proved for functions of random variable with regular varying right tails. For example, Proposition \ref{lemma:closure} B. implies the following: if $Z_1$,$Z_2$ are independent and have distributions with regularly varying right tails with coefficients $\alpha_1$ and $\alpha_2$, then $Z_1+Z_2$ will have a regularly varying right tail with coefficient $\min(\alpha_1,\alpha_2)$. The closure properties hold true for many different many-to-one functions such as the sum, maximum, product and ratio of random variables. The four classical problems described in \S 3 of our paper can be then recast as problems of regular variation on the prior $p(\psi)$, where the  marginal parameters of interest $\psi = \psi(\theta)$ satisfies the closure properties.

\section{Proofs of theorems}
\subsection{Proof of Theorem \ref{th:post-ss}}
\begin{proof}
Under the horseshoe prior, the prior on $\psi$ is proportional to 
$$
p( \psi ) \propto \int_{0}^{\infty} \psi^{p/2-1} \exp(-\psi/(2\tau^2)) \tau^{-p} (\tau^2+1)^{-1} d\tau.
$$
We can approximate the above integral using Laplace approximation as
\begin{align*}
p(\psi) & \propto \psi^{p/2-1} \int_0^{\infty} e^{-p\left(\frac{\psi}{2\tau^2 p}+\log \tau \right)} (\tau^2+1)^{-1} d\tau \\
& = \psi^{p/2-1} \int_0^{\infty} e^{M g(\tau)} h(\tau) d\tau,
\end{align*}
where $M = p$, $g(\tau) =  -\left(\psi (2\tau^2 p)^{-1}+\log \tau \right)$ and $h(\tau) = (\tau^2+1)^{-1} > 0$. The unique maximum of $g(\cdot)$ in $[0,\infty)$ is $\tau_0 = \sqrt{\psi/p}$, and $|g\rq{}\rq{}(\tau_0)|= 2p/\psi$. Hence, the Laplace approximation for the integral can be written as up to a proportionality constant independent of $\psi$ as
\begin{align*}
p(\psi) & \propto \psi^{p/2-1} \sqrt{\frac{2\pi}{M|g\rq{}\rq{}(\tau_0)|}} h(\tau_0) e^{M g(\tau_0)} \\
& \propto \psi^{p/2-1} \sqrt{\frac{2\pi}{|p/\psi|}} h(\tau_0) e^{-p(1/2+1/2 \log(\psi/p))} \\
& \propto \psi^{-\half} h(\psi/p) \propto \frac{1}{\sqrt{\psi} (\psi+p)}  .
\end{align*}
On the other hand, under the normal prior on $\theta$ the prior on $\psi$ is proportional to $p(\psi) \propto \psi^{ \frac{p}{2} -1 } e^{ - \frac{\psi}{2 \tau^2}  }$. Hence, noting the likelihood from Equation (4), the posteriors on $\psi$ when $Z\to \infty$ under the horseshoe prior and the gamma (induced by normal prior on $\theta_i$s) are given as: 
\begin{align*}
p_{HS}(\psi | Z) & \propto e^{-(\sqrt{z}-\sqrt{\psi})^2/2}\psi^{-(p+1)/4} \frac{1}{(\psi+p)} ,\\
p_{N}(\psi | Z) & \propto e^{-(\sqrt{z}-\sqrt{\psi})^2/2}\psi^{(p-3)/4} e^{ - \frac{\psi}{2 \tau^2}  }.
\end{align*}
\end{proof}

\subsection{Proof of Theorem \ref{th:horse-ss}}
\begin{proof}
\begin{align*}
P_{HS}(\psi \geq Z | Z) &\leq \frac{\int_{Z}^{\infty} e^{-(\sqrt{z}-\sqrt{\psi})^2/2}\psi^{-(p+1)/4} \frac{1}{(\psi+p)} d\psi }{\int_{0}^{\infty} e^{-(\sqrt{z}-\sqrt{\psi})^2/2}\psi^{-(p+1)/4} \frac{1}{(\psi+p)} d\psi} \\
& \leq \frac{\int_{Z}^{\infty} e^{-(\sqrt{z}-\sqrt{\psi})^2/2}\psi^{-(p+1)/4} \frac{1}{(\psi+p)} d\psi }{\int_{0}^{Z} e^{-(\sqrt{z}-\sqrt{\psi})^2/2}\psi^{-(p+1)/4} \frac{1}{(\psi+p)} d\psi}.
\end{align*}
Now, note that $\exp\{{-(\sqrt{z}-\sqrt{\psi})^2/2}\}$ is a decreasing function, which means it can be bounded by its values at the boundaries. Hence, 
\begin{align}
P_{HS}(\psi \geq Z | Z) & \leq \frac{\int_{Z}^{\infty} \psi^{-(p+1)/4} \frac{1}{(\psi+p)} d\psi }{\int_{0}^{Z} \psi^{-(p+1)/4} \frac{1}{(\psi+p)} d\psi} .\label{eq:tailprob}
\end{align}
Now, the integral in the denominator can be trivially bounded as: 
\beq
\int_{0}^{Z} \psi^{-(p+1)/4} \frac{1}{(\psi+p)} d\psi \geq Z^{-(p-3)/4} \frac{1}{Z+p} .\label{eq:lower}
\eeq
For upper bounding the numerator integral, we invoke Karamata's integral theorem, stated below. A proof can be found in \citet{bingham1989regular}.
\begin{theorem} (Karamata's theorem).
Let $L$ be a slowly varying and locally bounded function in $[x_0, \infty]$ for some $x_0 \geq 0$, then 
\ben 
\item for $\alpha > -1$, 
$$
\int_{x_0}^{x} t^{\alpha}L(t) dt \sim (\alpha+1)^{-1} x^{\alpha+1} L(x), \quad x \rightarrow \infty ,
$$
\item for $\alpha < -1 $, 
$$
\int_{x}^{\infty} t^{\alpha}L(t) dt \sim -(\alpha+1)^{-1} x^{\alpha+1} L(x), \quad x \rightarrow \infty .
$$
\een
\end{theorem}
Using the second part of Karamata\rq{}s integral theorem and the fact that $L(\psi) = \psi (\psi +p)^{-1}$ is a slowly varying function, we can derive an approximation for the numerator integral: 
\beq
\int_{Z}^{\infty} \psi^{-(p+1)/4-1} \frac{\psi}{(\psi+p)} d\psi \sim \frac{4}{p+1} Z^{-(p-3)/4} \frac{1}{Z+p} \quad \mbox{ as } Z \rightarrow \infty. \label{eq:upper}
\eeq
Putting \eqref{eq:lower} and \eqref{eq:upper} together, we can derive an upper bound for the tail probability in \eqref{eq:tailprob} as follows: 
\begin{align*}
P_{HS}(\psi \geq  Z | Z) & \leq \frac{4}{p+1} .
\end{align*}
\end{proof}


\subsection{Proof of Theorem \ref{th:norm-ss}}
\begin{proof}
To prove this assertion, we try to find a diminishing upper bound to the probability $P_{N}(\psi \leq Z | Z)$ using known inequalities. First, note as in Theorem~5.2, $e^{-(\sqrt{z}-\sqrt{\psi})^2/2}$ is a decreasing function and can be upper and lower bounded by its boundary values. 
\begin{align}
P_{N}(\psi \leq Z | Z) & \leq \frac{\int_{0}^{Z} e^{-(\sqrt{z}-\sqrt{\psi})^2/2}\psi^{(p-3)/4} e^{-\psi/2\tau^2} d\psi }{\int_{0}^{\infty} e^{-(\sqrt{z}-\sqrt{\psi})^2/2}\psi^{(p-3)/4} e^{-\psi/2\tau^2} d\psi} \nonumber\\
& \leq \frac{\int_{0}^{Z} \psi^{(p-3)/4} e^{-\psi/2\tau^2} d\psi }{\int_{0}^{\infty} \psi^{(p-3)/4} e^{-\psi/2\tau^2} d\psi} = \frac{1}{\Gamma(\frac{p+1}{4})} \gamma \left(\frac{p+1}{4}, \frac{Z}{2\tau^2} \right)\nonumber \\
&  \leq  \frac{1}{\Gamma(\frac{p+1}{4})} \frac{ \left(\frac{Z}{2\tau^2} \right)^{\frac{p+1}{4}} }{\frac{p+1}{4} } 
= \frac{1}{\Gamma(\frac{p+5}{4})} \left(\frac{Z}{2\tau^2} \right)^{\frac{p+1}{4}}, \label{eq:upper2}
\end{align}
where the last inequality follows from the well-known upper bound for the lower incomplete gamma function: 
\beq
\gamma(a,x) \leq \frac{x^{a-1}}{a}(1-e^{-x}) \leq \frac{x^{a}}{a} \quad \mbox{ for } x>0, a> 1 \nonumber. 
\eeq
Clearly, the upper bound in Equation \eqref{eq:upper2} goes to zero if $Z \tau^{-2} \rightarrow 0$. In  fact, the upper bound goes to zero if $Z \leq 2\tau^2$ and $p \rightarrow \infty$. 
\end{proof}

\section{Sampling Schemes for global-local priors}\label{sec:sampling}

Here we discuss two possible sampling strategies for the full Bayes inference under the horseshoe and  the horseshoe+ priors. 
\subsection{Exponential slice sampling for the horseshoe posterior}\label{sec:slice}
The horseshoe model is a hierarchical sparsity model.  We need to infer the location parameter $(\theta_1, \ldots , \theta_p)$
given data $y=(y_1, \ldots , y_p)$. Specifically, suppose that
\begin{eqnarray*}
(y_i \mid \theta_i ) & \sim & \Norm (\theta_i,1),\; (\theta_i \mid \lambda_i^2) \sim \Norm (0, \lambda_i^2 \tau^2), \\
\lambda_i   &\sim& C^+(0,1),\; \tau  \sim C^+(0,1),
\end{eqnarray*}
where $C^+(0,1)$ is a standard half-Cauchy distribution, so that: 
$$
p( \lambda_i  ) = \frac{2}{\pi} \frac{1}{ 1 + \lambda_i^2  } \; \; {\rm and} \; \; p(\tau) = \frac{2}{\pi} \frac{1}{1+\tau^2}.
$$
This can be re-parameterised in terms of shrinkage factors where $ \kappa_i = (1 + \lambda_i^2 \tau^2 )^{-1} $. We can marginalise out $\theta_i$ and write the joint posterior as:
$$
p( \kappa_1 , \ldots , \kappa_p , \tau^2 | y ) \propto
 \frac{\tau^p}{1+\tau^2} \prod_{i=1}^p \frac{ e^{- \kappa_i \frac{y_i^2}{2}}}{\sqrt{1-\kappa_i}} \frac{1}{\kappa_i(\tau^2-1) + 1}.
$$
Now introduce latent variables $ (\omega_1 , \ldots , \omega_p)$ and $\omega$ to write
\begin{align*}
\frac{1}{\kappa_i(\tau^2-1) + 1} & = \int_0^\infty  e^{- \omega_i (\kappa_i(\tau^2-1) + 1)} d \omega_i, \\
 \frac{1}{\tau^2+1} & =\int_0^\infty  e^{- \omega (\tau^2+1)} d \omega.
\end{align*}
Therefore, we have an augmented joint posterior: 
$$
p( \kappa_1 , \ldots , \kappa_p , \omega_1 , \ldots , \omega_p, \tau^2 , \omega | y ) \propto
 \tau^p e^{- \omega (\tau^2+1)}  e^{- \sum_{i=1}^p \kappa_i ( \omega_i (\tau^2-1) + y_i^2/2) - \omega_i} \prod_{i=1}^p \frac{1}{\sqrt{1-\kappa_i}} .
$$
Now introduce slice variables $(u_1, \ldots , u_p)$ on the set $\mathbb{I} \{0 < u_i < ( 1 -\kappa_i)^{-\frac{1}{2}} \}$. Therefore we have a joint posterior proportional to: 
$$
 \tau^p e^{- \omega (\tau^2+1)}  e^{- \sum_{i=1}^p \kappa_i ( \omega_i (\tau^2-1) + y_i^2/2) - \omega_i} \prod_{i=1}^p  
\mathbb{I}\{0 < u_i < ( 1 -\kappa_i)^{-\frac{1}{2}} \}.
$$
All the Gibbs conditionals are known in closed form.
$$
( \kappa_i | \omega_i , u_i,  \tau^2) \sim \mbox{Exp} \left (  \omega_i (\tau^2-1) + y_i^2/2 \right ) \mathbb{I} \{ 1 > \kappa_i > \max(1 - u_i^{-2},0) \},
$$
is a truncated exponential distribution.
$ ( \omega_i | \kappa_i , \tau^2 ) \sim \mbox{Exp} ( \kappa_i(\tau^2-1) + 1  ) $ and $ ( \omega | \tau^2 ) \sim \mbox{Exp} ( \tau^2 + 1 ) $ are exponentially distributed.
$ ( u_i| \kappa_i ) \sim \mbox{U} \{0 , ( 1 -\kappa_i)^{-\frac{1}{2}}\} $ follows a uniform distribution.
$ ( \tau^2 | \kappa , \omega ) \sim Ga \{ (p+1)/2 , \omega + \sum_{i=1}^p \kappa_i \omega_i \}$ follows a Gamma distribution. 
We can then simulate for the posterior $(\theta_i | y) $ using
$$
( \theta_i | \kappa_i , y_i ) \sim \Norm \left( ( 1 - \kappa_i)y_i , 1-\kappa_i \right).
$$

\subsection{Truncated normal slice sampling for the horseshoe+ posterior}\label{sec:trunc}
\citet{bhadra2015horseshoe+} show that for horseshoe+, the marginal prior on $\lambda_i$ is given by
$$
p( \lambda_i | \tau ) =   \left ( \frac{4}{\pi^2 \tau} \right ) \frac{ \ln \left \{ ( \lambda_i / \tau )^2 \right \}  }{ ( \lambda_i / \tau )^2 -1} 
\; \; {\rm where} \; \; \tau \sim p(\tau) \; .
$$ 
The posterior is proportional to
$$
p( \lambda_1 , \ldots , \lambda_p , \tau | y ) \propto \left \{ \prod_{i=1}^{p} \frac{1}{\sqrt{1+\lambda_i^2}} e^{-\frac{y_i^2}{2} \frac{1}{1+\lambda_i^2}} 
\frac{\log(\lambda_i/\tau)}{(\lambda_i/\tau)^2-1} \right \} \tau^{-p} p( \tau ).
$$
To simulate, we use the slice sampling and the Gibbs sampling. We introduce a vector of slice variables $ u =(u_1, \ldots , u_p)$ and simulate the joint distribution
$$
p( \lambda_1, \ldots , \lambda_p , u_1 , \ldots , u_p , \tau | y ) \propto 
\left \{ \prod_{i=1}^{p} \frac{1}{\sqrt{1+\lambda_i^2}} e^{-\frac{y_i^2}{2} \frac{1}{1+\lambda_i^2}} 
\mathbb{I}_{ \left\{ f(u_i) > \lambda_i/\tau \right\}} \right \} \tau^{-p} p(\tau) ,
$$
where $f^{-1}(x) = \log x / (x^2-1)$ is a monotonically decreasing function and so has a well defined inverse. If we marginalize out $ u$ then we have the required marginal. The complete conditional distributions are
\begin{align*}
p(\lambda_i | u_i,\tau, y) &\propto (1+\lambda_i^2)^{-1/2} \mbox{exp}[ -y_i^2/\{2(1+\lambda_i^2)\}] \mathbb{I}_{ \left\{ \lambda_i < \tau f(u_i) \right\}},\\
p(u_i | \lambda_i,\tau, y) &\propto \mathbb{I}_{\left\{ 0 < u_i < f^{-1}(\lambda_i/\tau) \right\}},\\
p(\tau | \lambda_i,u_i, y) &\propto \tau^{-p} p(\tau) \mathbb{I}_{ \left\{ \tau > \lambda_i / f(u_i) \; \forall i  \right\}}.
\end{align*}
The last conditional is $ p(\tau | \tau > \max ( \lambda_i / f(u_i) ) ) $. Posterior draws of $\theta$ can then be obtained by setting $\kappa_i = (1+ \lambda_i^2 \tau^2)^{-1}$ and 
$$
( \theta_i | \kappa_i , y_i ) \sim \Norm \left ( ( 1 - \kappa_i)y_i , 1-\kappa_i \right ).
$$

\section{MCMC Convergence Diagnostics}

\begin{figure}[ht!]%
\centering
\includegraphics[width = \linewidth,height=3in]{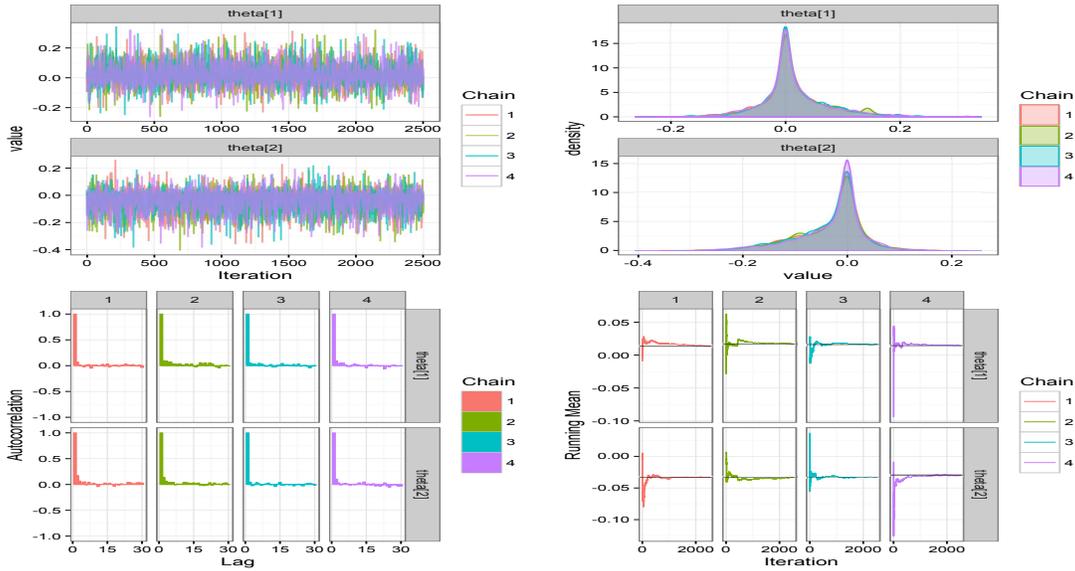}%
\caption{The convergence diagnostics plot for the bivariate examples under the horseshoe+ prior. The four panels, clockwise from top left, show the trace-plot, the density plot, the running mean plot and the auto-correlation plot.}%
\label{fig:diag}%
\end{figure}
We provide a few diagnostic plots that ensures the convergence of the Markov chain Monte Carlo sampling under the global-local shrinkage priors. For ease of illustration, we pick the bivariate case where the true values of the parameters $(\theta_1, \theta_2)$ is set at the origin $(0,0)$ and the horseshoe+ prior, as its convergence has not been studied elsewhere. We show four plots in Fig \ref{fig:diag} essential to judge convergence: the trace plot, the density plot, the running mean plot and the auto-correlation plot. The trace-plot shows the posterior samples as a time-series and the white-noise like plot in Fig \ref{fig:diag} ensures within-chain convergence or good mixing. The running mean plot shows the running mean of the sampling process shown in trace-plot and our plot suggests fast convergence to the target distribution. The density plot also shows agreement between the target distribution by different chains and finally the auto-corrleation plot in Fig \ref{fig:diag} shows low correlation between successive draws and good mixing. 

\section{Performance for Non-Gaussian Observations}

As suggested by a referee, we have investigated bivariate examples to see how well the the results extend beyond the Gaussian observation model. To this end, we simulated $K=100$ bivariate observations as independent draws from a $t$-distribution with $3$ degrees of freedom on each component of $(\theta_1, \theta_2)$. 
\begin{align*}
  (Y_{k,i} \mid \theta_i) &\sim t (d.f. = 3, n.c.p = \theta_i), i = 1,2, k = 1, \ldots, K. 
\end{align*}
We restrict ourselves to the case where the true values for both $\theta_1$ and $\theta_2$ are zero, that is, the observations are drawn from a central $t$-distribution with 3 degrees of freedom on each component. The sample mean for the data is $(\bar{y}_1, \bar{y}_2) = (0{\cdot}043,-0{\cdot}084)$. The results are given in  Fig. \ref{fig:t_fc_plot_theta_density} and Table \ref{tab:t-fc-and-pm}. It turns out that even in this case the global-local priors achieve stronger concentration near the origin compared to many of the other alternatives. 

\begin{figure}[ht!]%
\centering
\includegraphics[width = 0.8\linewidth,height  = 2.5in]{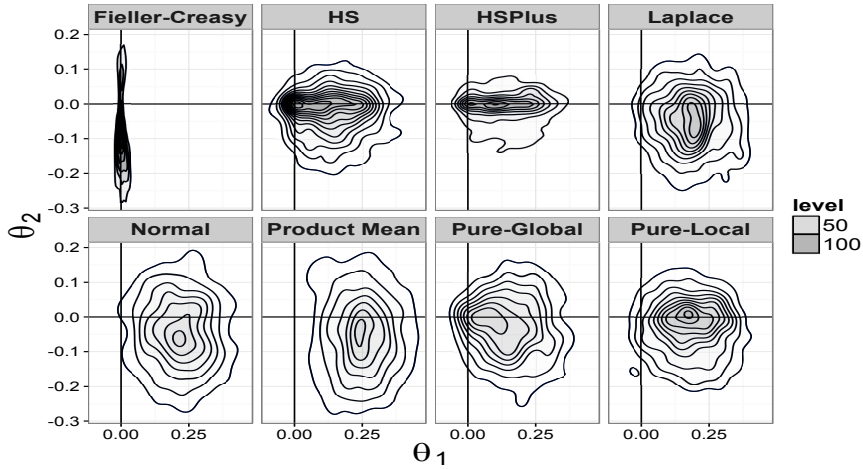}%
\caption{Two-dimensional contour plots of $p(\theta_1, \theta_2 \mid y)$ for the six shrinkage priors, namely, the horseshoe+ (HSPlus), the horseshoe (HS), Laplace, normal, pure-local, pure-global and the two reference priors for product mean and ratio of two means (Fieller-Creasy). }%
\label{fig:t_fc_plot_theta_density}%
\end{figure}

\begin{table}
\centering
\def~{\hphantom{0}}
\caption{Posterior quantiles for the marginal parameters of interest $\psi_1 = \theta_1/\theta_2$ (the Fieller--Creasy problem) and $\psi_2 = \theta_1 \theta_2$ for the candidate priors. All entries in the table have been multiplied by $1,000$ for the sake of representation.} {%
{\footnotesize
\begin{tabular}{rrrrrrrr}
       & \multicolumn{3}{c}{$\psi = \theta_1/\theta_2$} &       & \multicolumn{3}{c}{$\psi = \theta_1 \theta_2 $} \\
      & $Q_1$ & Median  & $Q_3$ &       & $Q_1$ & Median  & $Q_3$ \\
HS+   & -5010 & -628${\cdot}$8 & 3776  & HS+   & -4${\cdot}$7  & -0${\cdot}$2  & 1${\cdot}$2 \\
HS    & -3450 & -739${\cdot}$3 & 2273  & HS    & -7${\cdot}$1  & -0${\cdot}$5  & 1${\cdot}$5 \\
Laplace & -3057 & -1316 & 1270  & Laplace & -17${\cdot}$9 & -5${\cdot}$5  & 1${\cdot}$4 \\
Normal & -3061 & -1219 & 1928  & Normal & -24${\cdot}$8 & -6${\cdot}$7  & 5${\cdot}$1 \\
Local & -3541 & -1007 & 2662  & Local & -11${\cdot}$9 & -1${\cdot}$6  & 3${\cdot}$0 \\
Global & -2669 & -875${\cdot}$8 & 1816  & Global & -13${\cdot}$3 & -1${\cdot}$7  & 3${\cdot}$1 \\
Reference & -89${\cdot}$5 & -18${\cdot}$6 & 53${\cdot}$8 & Reference & -30${\cdot}$4 & -10${\cdot}$0  & 5${\cdot}$4 \\
\end{tabular}%
}%
}
\label{tab:t-fc-and-pm}
\end{table}

\section{Hyper-parameter for for $\tau$}
We performed a prior sensitivity analysis to justify the choice of hyper-parameter $\eta=1$ for the scale of the half-Cauchy prior on $\tau$ in the hierarchical model for the horseshoe prior: 
\begin{align*}
  ( \theta_i \mid \lambda_i, \tau ) &\sim \Nor \left ( 0 , \lambda_i^2 \right ), ( \lambda_i \mid \tau ) \sim C^+ \left ( 0 , \tau \right ), \tau  \sim C^+ \left ( 0 , \eta \right ), \lambda_i, \tau > 0, \eta>0.
\end{align*}
To show that the hyper-parameter $\eta$ has minimal effect on inference, we study the performance of the horseshoe prior for the Efron's sum-of-squares problem with a sparse $\theta$ with a single non-zero component of magnitude $10$ and compare the posterior summary for three different values of $\eta \in \{0.5,1,5\}$. The observed value of $\sum_{i=1}^{n} y_i^2$ is $199{\cdot}3$. As Table \ref{tab:compare-tau-summary} and Fig. \ref{fig:compare-tau-density} show, the posterior density under the three different choices of scale parameters are very similar, as are the posterior mean and quantiles, with the closest answer being given by the choice of $\eta = 1$. This supports our choice of unit scale for the global-shrinkage parameter $\tau$. 

\begin{figure}[ht!]%
\centering
\includegraphics[width = 0.5\linewidth, height = 2in]{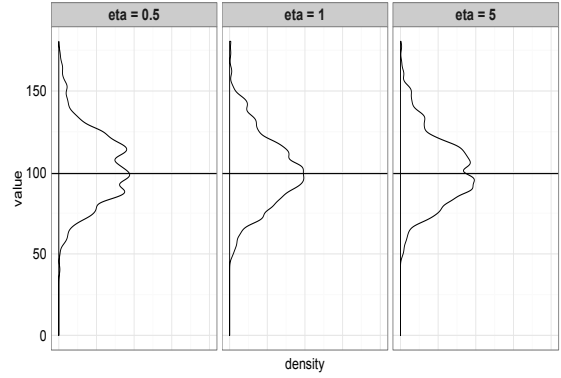}%
\caption{Posterior densities of the parameter of interest $\psi = \sum_{i=1}^{p} \theta_i^2$, for the horseshoe prior with $\eta\in\{0.5, 1, 10\}$. The horizontal is at the true value $\psi = 100$.  }%
\label{fig:compare-tau-density}%
\end{figure}

\begin{table}[!h]
\centering
\def~{\hphantom{0}}
\caption{Summary statistics and standard deviation of the posterior distribution under the horseshoe prior for $\eta \in \{0.5,1,5\}$ for the sum-of-squares problem. The true $\theta$ is a $100$-dimensional sparse vector with one non-zero component with magnitude $\theta_1 = 10$ and $\sum_{i=1}^{100} y_i^2 = 199.3$.}{%
{\footnotesize
\begin{tabular}{rrrrrrrr}
      & Minimum  & $Q_1$ & Median  & Mean  & $Q_3$ & Max   & Std. Dev.  \\
$\eta$ = 0${\cdot}$5 & 40${\cdot}$38 & 88${\cdot}$3  & 102${\cdot}$2 & 103${\cdot}$5 & 117${\cdot}$2 & 173${\cdot}$1 & 21${\cdot}$4 \\
$\eta = 1$ & 48${\cdot}$64 & 87${\cdot}$67 & 100${\cdot}$7 & 101${\cdot}$8 & 115${\cdot}$1 & 180${\cdot}$5 & 21${\cdot}$4 \\
$\eta = 5$ & 49${\cdot}$45 & 88${\cdot}$3  & 101${\cdot}$1 & 102${\cdot}$5 & 115${\cdot}$6 & 177${\cdot}$3 & 21${\cdot}$1 \\
\end{tabular}%
}%
}
\label{tab:compare-tau-summary}
\end{table}

\section{Connections with the Penalized Complexity Priors}

The penalized complexity (PC) prior of \citet{simpson2014pc} on the precision parameter $\tau$ is given in Equation (3.3) of their article
$$
p(\tau) = \frac{\lambda}{2} \tau^{-3/2} \exp(-\lambda \tau^{-1/2}), \tau >0,
$$
where the parameter $\theta$ is a normal scale mixture, defined as, $\theta\mid\tau \sim \Nor (0, \tau^{-1} R^{-1})$ for some positive semi-definite $R$ and the data distribution can be taken as $y\mid \theta\sim  \Nor(0, \sigma^2\mathrm{I})$. Without loss of generality, let $R$ be the identity matrix. A change of variables $\sigma^2 = \tau^{-1}$ yields the prior on the variance
$$
p(\sigma^2) = \frac{\lambda}{2} (\sigma^2)^{-1/2} \exp\{-\lambda (\sigma^2)^{1/2}\}, \sigma^2 >0.
$$
Thus, $p(\sigma^2) \sim (\sigma^2)^{-1/2}$ when $\sigma^2$ is close to zero and $p(\sigma^2) \sim \exp(-\lambda (\sigma^2)^{1/2})$ when $\sigma^2$ is large. This shows the PC prior on $\sigma^2$ has an unbounded mass at the origin, and Reviewer 1 is correct that this is an improvement over the inverse gamma prior on $\sigma^2$. However, the prior on $\sigma^2$ still has an exponentially decaying tail. As a result, the marginal PC prior on $\theta$ should behave similarly to the horseshoe prior on $\theta$ around the origin. It will, however, shrink the large signals due to the exponential tails of $\sigma^2$. Contrast this to the horseshoe priors on $\theta$ that results in both an unbounded mass at the origin as well as polynomially decaying tails for $\theta$ \citep[Fig.~1 of][]{carvalho2010horseshoe}. This distinction may not matter much in hypothesis testing problems, where separating zero versus non-zero is the major aim, but an application of Theorem 2 of \citet{carvalho2010horseshoe} shows the performance of PC priors in estimating large signals will be sub-optimal. In particular, the regularly-varying tails of the horseshoe priors on the variance parameter that results in the regularly varying tails of the marginals of $\theta$ (see Theorem 2.1 of the revised manuscript) does not appear to be applicable to the PC priors.

\citet{simpson2014pc} admit to this behavior of the PC prior in several places of their manuscript. On page 18, after Theorem 3 they say ``This result suggests the PC prior will shrink strongly, \ldots, due to the relatively light tail of the exponential.'' They also devote Section 4.5 of the article to point out the unsuitability of the PC priors as sparsity priors and concludes the section by noting ``This is the only situation we encountered in which the exponential tails of the PC priors are problematic.'' Thus, although we have not compared it's simulations, it appears the PC priors were designed with a different goal in mind from the problems we consider here.
}
\bibliographystyle{biometrika}
\bibliography{horseshoe-plus}
\end{document}